\documentclass[12pt]{article}

\usepackage{tablefootnote}

\usepackage{amsthm}

\usepackage[titletoc]{appendix}

\renewcommand{\baselinestretch}{1.4}

\newcommand{\filtration}{{\mbox{\boldmath$\mathcal{F}_{t-1}$}}}

\usepackage{subfigure,float}
\usepackage[default]{jasa_harvard}    

\usepackage{amsmath,amssymb}
\usepackage{color}
\usepackage{latexsym}
\usepackage{graphicx,graphics}
\DeclareGraphicsExtensions{.bmp,.jpg,.eps,.pdf}
\usepackage{float,verbatim}
\begin{document}

\title{A New Class of Discrete-time Stochastic Volatility Model with Correlated Errors}

\renewcommand{\baselinestretch}{1.0}
\author{Sujay Mukhoti and Pritam Ranjan\\[0.1in]
Operations Management and Quantitative Techniques,\\
Indian Institute of Management Indore, M.P., India, 453556\\
(sujaym@iimidr.ac.in, pritamr@iimidr.ac.in)
}
\date{}

\maketitle

\begin{abstract}
In an efficient stock market, the returns and their time-dependent volatility are often jointly modeled by  stochastic volatility models (SVMs). Over the last few decades several SVMs have been proposed to adequately capture the defining features of the relationship between the return and its volatility. Among one of the earliest SVM, \citeasnoun{Taylor1982} proposed a hierarchical model, where the current return is a function of the current latent volatility, which is further modeled as an auto-regressive process. In an attempt to make the SVMs more appropriate for complex realistic market behavior, a leverage parameter was introduced in the Taylor's SVM, which however led to the violation of the efficient market hypothesis (EMH, a necessary mean-zero condition for the return distribution that prevents arbitrage possibilities). Subsequently, a host of alternative SVMs had been developed and are currently in use. In this paper, we propose mean-corrections for several generalizations of Taylor's SVM that capture the complex market behavior as well as satisfy EMH. We also establish a few theoretical results to characterize the key desirable features of these models, and present comparison with other popular competitors. Furthermore, four real-life examples (Oil price, CITI bank stock price, Euro-USD rate, and S\&P 500 index returns) have been used to demonstrate the performance of this new class of SVMs. \\

\end{abstract}
\renewcommand{\baselinestretch}{1.5}

\section{Introduction}\label{sec:intro}

Over the past few decades, time-varying volatility of asset returns has drawn significant attention in financial statistics. The volatility of asset return is often defined as the standard deviation or variance of the returns and is assumed to be unobservable. For the theoretical and empirical results presented in this paper, we use the variance of the returns as the volatility. A financial market is said to be efficient if the price of a risky asset (e.g. equity or stock) contains every available information about it, which is referred to as \emph{efficient market hypothesis} (EMH). In such a market the risk of an asset is measured by its return \emph{volatility}. Stochastic volatility models (SVMs) represent a popular class of hierarchical models for describing the relationship between asset return and its time-varying volatility. Let $\varepsilon_t$ and $\eta_t$ denote the errors in the return and log-volatility process, then this paper focusses on the SVMs with correlated return-volatility relationship, i.e., $\rho = Corr(\varepsilon_t, \eta_t) \ne 0$. Although the concept of correlated return and volatility in continuous-time SVM dates back to \citeasnoun{Black1976177}, the discrete-time correlated SVMs have not been investigated much until recently. In this paper, we also study the pattern of correlations between current return and lagged or lead volatilities and propose a set of new discrete-time SVMs..

Suppose $P_t$ denotes the price of a risky asset at time $t$, then the mean-adjusted return $r_t = \log(P_t / P_{t-1})$, can be modeled using an SVM. Although there is a plethora of SVMs for describing the returns, one of the simplest yet most popular discrete-time SVM is given by \citeasnoun{Taylor1982}, where the return process $r_t$ is a non-linear product of two independent stochastic processes, viz. an i.i.d. error process $\varepsilon_t$, and a latent log-volatility process $h_t$, which is further modeled as an $AR(1)$. That is,
\begin{eqnarray}
  r_t & = & \exp\left\{\frac{h_{t}}{2}\right\}\varepsilon_t, \nonumber \\
  h_t & = & \alpha +\phi(h_{t-1}-\alpha)+\sigma \eta_t, \; \forall t=1, 2, \ldots ,
  \label{eq:SVM_taylor}
\end{eqnarray} 
where $\alpha = E(h_t)$ is the long-range log-volatility, $\phi$ captures the stationarity of the log-volatility process, $\sigma$ measures the variability of $h_t$, and $\varepsilon_t$ and $\eta_t$ are uncorrelated i.i.d. $N(0, 1)$ errors. Notice that in this case $r_t$ is a martingale difference sequence so that $E[r_t \mid \filtration]=0$, where $\mathcal{F}_{t-1}$ is the space ($\sigma$-field) generated with $r_1, ..., r_{t-1}$. In other words, the return is not predictable by past observations and hence comply with EMH.

\citeasnoun{Black1976177} pointed out that high volatility is coupled with price drop (or negative return), and low volatility follows price increase (or positive return). This negative correlation between return and its volatility is termed as ``leverage" effect (see e.g., \citeasnoun{Nelson1991}). Among others, \citeasnoun{Jacquier2004185} suggested using a correlation parameter $\rho = Corr(\varepsilon_t, \eta_t)$ to capture more realistic market behaviour. However, it turns out that a non-zero $\rho$ parameter makes $E[r_t \mid \filtration ]$ non-zero, which violates EMH \cite{Yu2005165}. This led to a roadblock for further extensions and generalizations of model (\ref{eq:SVM_taylor}).

Alternatively, the relationship between the return and log-volatility can be modelled as
\begin{eqnarray}
		r_t&=&\exp\left\{\frac{h_t}{2}\right\}\varepsilon_t, \nonumber \\
		h_{t+1}&=&\alpha+\phi(h_t-\alpha)+\sigma\eta_t   \label{eq:htplus1_base_model}
	\end{eqnarray}
where the correlation between $\eta_t$ and $\varepsilon_t$ is $\rho$, and in the second level of the hierarchical structure, $h_{t+1}$ is a function of $(h_{t}, \eta_t)$, as compared to model (\ref{eq:SVM_taylor}), where $h_t$ is modelled with respect to $(h_{t-1}, \eta_t)$ in the AR(1) structure (see \citeasnoun{Ghysels1996119} and \citeasnoun{Omori2007425} for details). Since in model (\ref{eq:htplus1_base_model}), $h_t$ depends on $\eta_{t-1}, \eta_{t-2}, ...,$ it is straightforward to show that $E[r_t\mid \filtration]=0$, which ensures concordance with EMH.

A host of generalizations of this SVM, model (\ref{eq:htplus1_base_model}), have been proposed in the past few years to make the model more realistic that can capture complex features of return processes. For instance, \citeasnoun{Duffie2000} and \citeasnoun{Eraker2003} used jump components in the return and volatility processes to capture extreme returns and their persistent effects caused by crash-like events which are  not too rare. \citeasnoun{Aas2006275} used generalized hyperbolic skewed-$t$ distribution to explicitly account for the skewness and heavy-tails in the return distribution; \citeasnoun{Abanto-Valle20102883} used scale mixture of normals; and \citeasnoun{Dipak2015} used skewed-$t$ distribution for modelling the returns.

We categorize the SVMs in two classes, (a) $h_{t+1}$-based models - generalizations of model (\ref{eq:htplus1_base_model}) and (b) $h_t$-based models - generalizations of model (\ref{eq:SVM_taylor}). Despite the abundance of $h_{t+1}$-based generalizations, not many $h_t$-based SVMs have been developed thus far. As per our understanding, the main reason behind the scarce of $h_t$-based generalizations is the failure of (\ref{eq:SVM_taylor}) with correlated errors in satisfying EMH. In an attempt to address this issue, \citeasnoun{Mukhoti2016} suggested a mean-corrected version of (\ref{eq:SVM_taylor}) with correlated errors.

The main focus of this paper is to extend the work of \citeasnoun{Mukhoti2016} and develop generalized $h_t$-based SVMs corresponding to the generalized $h_{t+1}$-based models  with skewed-$t$ errors \cite{Dipak2015} and jump components \cite{Eraker2003}. Furthermore,  \citeasnoun{Mukhoti2016} defined the lead-lag correlation as $Corr(r_t, h_{t\pm k})$, whereas, in this paper, we follow the more conventional approach and use $Corr(r_t, e^{h_{t\pm k}})$ for quantifying lead-lag correlations (note that $h_t$ is the log-volatility and $e^{h_t}$ is the volatility).

For daily frequency data, \citeasnoun{Bollerslev2006} reports (model-free) correlation between $r_t$ and a proxy of volatility, $r_{t\pm k}^2$ (since $h_t$ is unobservable), and demonstrate that the contemporaneous correlation between return ($r_t$) and its volatility ($e^{h_t}$) is negative and maximum in magnitude, and $Corr(r_t, e^{h_{t+k}})$ increases exponentially towards zero  with respect to $k>0$. \citeasnoun{ait2013} provides the estimation biases of the contemporaneous return-volatility correlation when volatility proxy is the realized volatility. These two papers establish the contemporaneous correlation as the leverage effect. We have not come across any other research on leverage in the light of different SVM specifications, i.e., model implied leverage. In this paper we derive the leverage (or contemporaneous correlation) using both approaches $Corr(r_t, r_t^2)$ as in \citeasnoun{Bollerslev2006} and model implied $Corr(r_t,e^{h_t})$. We generalize this further to find lead-lag correlations  between $r_t$ and $e^{h_{t\pm k}}$ for $k>0$.

We also present closed form expressions for the first four unconditional moments of the return distribution. These moments are further used for computing skewness and kurtosis that quantifies the desired asymmetry and tail-fatness in the return distribution. Though the derivations (for third and fourth order moments and lead-lag correlations) are not too tricky, we are not aware of its existence in the literature. These summary statistics facilitate comparison of the proposed $h_{t}$-based models with the corresponding $h_{t+1}$-based SVMs. Considering the length of the manuscript, the derivations and proofs have not been included here. In addition to the theoretical comparison, we fit the two classes of models on four real-life datasets, viz. daily oil reference basket (ORB) price, CITI bank stock price, S\&P 500 index and Euro-US dollar exchange rates, and compare different model features. The datasets are chosen specifically to represent different types of risky assets. For model implementation, we used the Bayesian framework under \emph{Just Another Gibbs Sampler} (JAGS) software.

The rest of the manuscript is organized as follows: Section 2 presents a brief review of three popular $h_{t+1}$-based SVMs (base model (\ref{eq:htplus1_base_model}), model with skewed-$t$ return, and model with jump components). Section~3 starts with a brief recap of the (base model) results in \citeasnoun{Mukhoti2016}, then we present new mean-corrected $h_t$-based models with skewed-$t$ return distribution and jump components. Section~4 outlines a few theoretical results on moments and lead-lag correlations that compare key features of the two classes ($h_t$- and $h_{t+1}$-based) of SVMs. Section~5 summaries the implementation results of the six models on four real-life applications, and Section~6 concludes with a few important remarks.

\section{Popular $h_{t+1}$-based SVMs}\label{sec:pophtplus1SVM}

In this section, we briefly review three popular $h_{t+1}$-based SVMs {with correlated errors} $\varepsilon_t$ and $\eta_t$. The closed form expressions of the first four moments of $r_t$, and lead-lag correlations between $r_t$ and $e^{h_{t\pm k}}$ under these models are presented in Section~4. The derivations are not tricky, however, to the best of our knowledge, the closed form expressions for the third and fourth order moments and lead-lag correlations are not explicitly available in the literature.

\textbf{(M2.1) Base model}: Though the first discrete-time SVM was formally proposed by \citeasnoun{Taylor1982}, as shown in (\ref{eq:SVM_taylor}), \citeasnoun{Ghysels1996119} documents one of the simplest yet realistic $h_{t+1}$-based SVM that can capture leverage effect and the volatility clustering in return-volatility relationship. We refer to (\ref{eq:htplus1_base_model}) as the base model in this class of SVMs. In this model, we assume that $(\varepsilon_t, \eta_t)$ follows a bi-variate normal distribution with mean zero, variance $1$ and correlation $\rho$. The unconditional central moments of $r_t$ (presented in Section~4) can be used to measure skewness and kurtosis. Furthermore, since $h_{t+1}$ can be expressed as
$$ h_{t+1} - \alpha = \sigma \sum_{j=0}^{\infty} \phi^{j} \eta_{t-j},$$
the marginal distribution of $h_t$ is $N\left(\alpha, \sigma^2/(1-\phi^2)\right)$.

This model, (\ref{eq:htplus1_base_model}), has been a basis of several applications and methodological research in the recent literature. For instance, \citeasnoun{Omori2007425} used Markov Chain Monte Carlo (MCMC) based parameter estimation technique with this model for Tokyo stock exchange daily returns. A multivariate extension of this basic SVM was discussed by \citeasnoun{Asai2009292}. \citeasnoun{Du2011497} used the above model to explain the crude oil return-volatility relationship. \citeasnoun{Yu2012473} proposed a semi-parametric generalization of the classical leverage structure present in this SVM, where the efficiency of the model was demonstrated using S\&P500 index and Microsoft stock returns data observed in daily and weekly frequency.

The bi-variate normal distribution of  $(\varepsilon_t, \eta_t)$ used in M2.1 falls inadequate to explain extreme returns, which occasionally appears in many real-life situations, for example, {in a crash like period such as 2008 Lehman Brothers incident}. As a result, further update / extension of (\ref{eq:htplus1_base_model}) becomes necessary. A natural alternative is to use a heavy-tail distribution instead of Gaussian for modelling the returns. \citeasnoun{Chib2002281} generalized M2.1 by modelling returns with a $t$-distribution, i.e., $r_t = \omega\exp\left(h_t/2\right)U_t^{-1/2} \varepsilon_t$, where $U_t \sim \Gamma(\nu/2, \nu/2)$, $\varepsilon_t / \sqrt{U_t}$ follows a $t$-distribution with $\nu$ degrees of freedom, and $\omega$ is chosen such that $Var(r_t|h_t)=\exp(h_t)$.  \citeasnoun{Berg2004107} and \citeasnoun{Asai2008332} compared several SVMs with different heavy-tailed return distributions. \citeasnoun{Abanto-Valle20102883} proposed a robust Bayesian SVM for fat tails of returns using scale mixture of normals. \citeasnoun{Wang2011852} modelled $(\varepsilon_t, \eta_t)$ as a bi-variate-$t$ distribution, which might be questionable, as the moment generating function of $\eta_t$ and subsequently any marginal moment of $r_t$ does not exist. However, this is not a concern if we are interested in (say) the conditional distribution of $[r_t|h_t]$ and not the marginal.

Of course, a necessary assumption of zero mean return ($E[r_t\mid\mathcal{F}_{t-1}]=0$, as per EMH) for any acceptable SVM, does not guarantee the return distribution to be symmetric. In fact, as indicated by  \citeasnoun{Black1976177}, $Var(r_t|r_t>0) < Var(r_t|r_t<0)$, which induces asymmetry in the return distribution. As a result, the typical Gaussian or $t$ distributions as presented in such SVMs are not capable of capturing this asymmetry. The third moment of $r_t$-distribution and hence its skewness under M2.1 are also zero (see Table~\ref{tab:SVmoments} in Section~4). \citeasnoun{Tsiotas2011} developed an SVM with leverage and skewed-$t$ errors, but it does not comply with EMH (i.e., $E[r_t\mid\mathcal{F}_{t-1}] \ne 0$), and \citeasnoun{Dipak2015} proposed a skewed-$t$ based model but without the correlated errors. We next present a natural generalization of the SVM by \citeasnoun{Dipak2015}, and include correlated return and volatility errors.

\textbf{(M2.2) Skewed $t$ model}: The returns can be modelled as 
\begin{eqnarray}\label{eq:m2.4}
 r_t &=& \exp\left(\frac{h_t}{2}\right) \omega U_t^{-1/2}\left[\delta \left(W_t-\sqrt{\frac{2}{\pi}} \right) +  \sqrt{1-\delta^2}\varepsilon_t\right] \\ \nonumber
 h_{t+1} &=& \alpha + \phi(h_{t} - \alpha) + \sigma\eta_t,
\end{eqnarray}
where $(\varepsilon_t, \eta_t)$ follows a bi-variate normal with mean zero, variance $1$ and correlation $\rho$,  the weight constant is $\delta = \lambda/\sqrt{1+\lambda^2}$, $U_t \sim \Gamma(\nu/2, \nu/2)$, $W_t \sim N_+(0,1)$ (half normal), and $\omega$ is chosen such that $Var(r_t|h_t)=\exp(h_t)$. Here, $S_t = \omega U_t^{-1/2}\left[\delta \left(W_t-\sqrt{2/\pi} \right) +  \sqrt{1-\delta^2}\varepsilon_t\right]$ follows a skewed-$t$ distribution, with skewness defined by $(W_t-\sqrt{2/\pi})/\sqrt{U_t}$, and $\varepsilon_t/\sqrt{U_t}$ accounts for the heavy-tail part via a $t$-distribution. As earlier, the marginal distribution of $h_t$ is $N\left(\alpha, \sigma^2/(1-\phi^2)\right)$, and the unconditional central moments of $r_t$ are summarized in Section~4.

Alternatively,  \citeasnoun{Aas2006275}, \citeasnoun{Nakajima20123690} and \citeasnoun{Takahashi2016437} have modeled the tail-fatness and skewness in returns using a generalized hyperbolic skewed-$t$ distributions, \citeasnoun{Abanto-Valle20102883} used scale mixture of normals, whereas, \citeasnoun{delatola2011} and \citeasnoun{Jensen2010306} used Dirichlet process mixture to capture the return skewness and kurtosis. Note that none of the aforementioned SVMs can capture sudden drop (or rise) in returns observed during crashes (or boom). Such (non-stationary) changes in the returns are better explained by jumps in the return distribution \cite{Nakajima20092335}.

\textbf{(M2.3) Model with jump}: \citeasnoun{Eraker2003} present one of the most popular SVM with jump components included in both return and log-volatility processes to accommodate high volatility required to generate such extreme returns. The model is given as follows:
\begin{eqnarray}\label{eq:m2.2}
 r_t &=& K_{1t}J_{1t} + \exp\left(\frac{h_t}{2}\right) \varepsilon_t,  \nonumber \\
 h_{t+1} &=& K_{2t}J_{2t} + \alpha + \phi(h_{t} - \alpha) + \sigma\eta_t,
\end{eqnarray}
where $(\varepsilon_t, \eta_t)$ follows the same bi-variate normal distribution with mean zero, variance $1$ and correlation $\rho$, the jump coefficients $K_{it}\sim N(\nu_i, \tau^2_i)$, the jump components $J_{it} \sim Ber(\pi_i)$, and $K_{it}, J_{it}$ are all independent with each other and with $\varepsilon_t$ and $\eta_t$. The compliance with EMH requires $E[K_{1t}]=0$, that is, $\nu_1=0$. For simplicity, we also assume that $E[K_{2t}]=\nu_2=0$ and $\tau_1=\tau_2=1$ for all the results in this paper, however, the results can easily be extended for general $\nu_2$, $\tau_1$ and $\tau_2$. The jump component in the log-volatility process affects the marginal distributions of $h_t$, and it is no longer a normal, however, its long-term average volatility is still $\alpha$ and variance becomes $(\sigma^2+\pi_2(1-\pi_2))/(1-\phi^2)$. Table~\ref{tab:SVmoments} summarizes the central moments of the marginal distribution of $[r_t|\filtration]$.

Indeed the jump components are capable of capturing transient price changes in the sense that the distribution of the future returns are not influenced by the currently observed extreme value. Recently, \citeasnoun{Liu2014} developed a Bayesian unit root test for the above SVM with jumps.

Undoubtedly, the evolution of $h_{t+1}$-based SVMs over the last few years has been extensive, however, there are several fundamental aspects of SVMs that require further investigation. For instance, as shown in Table~\ref{tab:SVmoments}, the third central moments $E(r_t^3)$, and hence the skewness measured by $E(r_t^3) / (Var(r_t)^{3/2})$, is zero for M2.1 and M2.3 and not M2.2. This may seem natural as only M2.2 incorporates a skewed-$t$ component, however, it is not in concordance with the recent findings of strong instantaneous leverage (see, \citeasnoun{ait2013} and \citeasnoun{wang2014}, among others). Moreover, $Cov(r_t, r_t^2) = E(r_t^3) = 0$ contradicts the basis of defining leverage as contemporaneous correlation suggested by \citeasnoun{Bollerslev2006}. Next we present a new class of $h_t$-based SVMs with correlated errors.

\section{New Class of $h_t$-based SVMs} \label{sec:newSVM}

In this section, we propose a modification in the $h_{t+1}$-based SVMs, M2.1, M2.2 and M2.3, presented in Section~2. The main idea is to change the second stage of the hierarchical model from ``$h_{t+1}$ as a function of $(h_t, \eta_t)$'' to ``$h_t$ as a function of $(h_{t-1}, \eta_t)$''. However, this modification induces a nonzero marginal expected return $E(r_t|\filtration)$, which is unacceptable as the EMH assumption is violated, and may lead to arbitrage opportunities. Consequently, we also propose a mean-correction in the first stage of the hierarchical model structure to ensure $E(r_t|\filtration)=0$.  This idea is inspired from \citeasnoun{Jacquier2004185}, where the authors modified (\ref{eq:SVM_taylor}), the model by \citeasnoun{Taylor1982}, however, as pointed out by \citeasnoun{Yu2005165}, the proposed modification also violated EMH. Recently, \citeasnoun{Mukhoti2016} developed a correctly modified $h_t$-based base model. In this paper, we present this modified SVM as M3.1, and present a few additional interesting properties. We also extend \citeasnoun{Mukhoti2016}, and develop two new $h_t$-based models corresponding to M2.2 and M2.3 which are referred to as M3.2 and M3.3, respectively.

\textbf{(M3.1) Base model}: \citeasnoun{Mukhoti2016} suggested the following mean-corrected SVM as the simplest yet generalized model with desirable  properties:
\begin{eqnarray}\label{eq:m3.1}
 r_t &=& \mu_{1} + \exp\left(\frac{h_t}{2}\right) \varepsilon_t, \\ \nonumber
 h_{t} &=& \alpha + \phi(h_{t-1} - \alpha) + \sigma\eta_t,
\end{eqnarray}
where $(\varepsilon_t, \eta_t)$ follows a bi-variate normal with mean zero, variance $1$ and correlation $\rho$. \citeasnoun{Mukhoti2016} show that under the regularity condition of  $|\phi|\le 1, \sigma>0$ and $-\infty< \alpha <\infty$, the mean correction term is
$$	\mu_1 = - \frac{\rho\sigma}{2}\exp\left\{\frac{\alpha}{2}+\frac{\sigma^2 }{8(1-\phi^2)}\right\}, 	$$
and the higher order moments of the unconditional marginal distribution of $r_t$ is summarized in Table~\ref{tab:SVmoments} of Section~4 of this paper. Similar to the corresponding $h_{t+1}$-based model (M2.1), the marginal distribution of $h_t$ is $N\left(\alpha, \sigma^2/(1-\phi^2)\right)$. 

Contrary to M2.1, even the normal distribution of errors ($\eta_t, \varepsilon_t$) give non-zero third order central moment of $r_t$ , i.e., $r_t$ distribution in (\ref{eq:m3.1}) can capture some amount of skewness and in-turn capable of explaining leverage to an extent. However, only the contemporaneously correlated errors, $Corr(\eta_t, \varepsilon_t)=\rho$, may not suffice to account for the asymmetric return-volatility relationship (see e.g., \citeasnoun{figlewski2001}). This necessitates development of a new model with contemporaneously correlated errors as well as (marginally) skewed returns.

\textbf{(M3.2) Skewed $t$ model}: As in M2.2, this model enforces the skewed-$t$ distribution of returns to capture the heavy-tails and asymmetry. The model is given by 
\begin{eqnarray}\label{eq:m3.4}
 r_t &=& \mu_2 + \exp\left(\frac{h_t}{2}\right) \omega U_t^{-1/2}\left[\delta \left(W_t-\sqrt{\frac{2}{\pi}}\right) +  \sqrt{1-\delta^2}\varepsilon_t\right], \\ \nonumber
 h_{t} &=& \alpha + \phi(h_{t-1} - \alpha) + \sigma\eta_t,
\end{eqnarray}
where $(\varepsilon_t, \eta_t)$ follows the same bi-variate normal with mean zero, variance $1$ and correlation $\rho$, the {constant $\omega$ is } such that $Var(r_t|h_t)=\exp(h_t)$, $\delta = \lambda/\sqrt{1+\lambda^2}$, $U_t \sim \Gamma(\nu/2, \nu/2)$ and $W_t \sim N_+(0,1)$ (half normal). Note that the additional mean-correction term which enables EMH is given by
$$
	\mu_2=-\frac{\rho\sigma}{2}\sqrt{1-\delta^2}\cdot \omega \xi_\nu\left( \frac{1}{2} \right) \exp\left\{\frac{\alpha}{2}+\frac{\sigma^2}{8(1-\phi^2)}\right\},
$$
where $\xi_\nu(k)=E\left[ U_t^{-k} \right]$. As in M2.2 and M3.1, here also, the marginal distribution of $h_t$ is $N(\alpha, \sigma^2/(1-\phi^2))$. See Table~\ref{tab:SVmoments} for the first four central moments of the  marginal distribution of $r_t$.

The special case of $\delta=0$ simplifies M3.2 and focusses only on the heavy-tail component. Despite forcing $\delta=0$, the third moment is not identically zero, and as in the base model M3.1, the simplified model explains the skewness to some extent. Of course, the general case contains additional terms that specifically accounts for the skewness (see Table~1).

\textbf{(M3.3) Model with jump}: 
We now propose the mean corrected $h_t$-based version of M2.3. This new model is capable of (i) explaining the non-zero contemporaneous correlation and return skewness, and (ii) generating extreme return followed by similar values during immediate next periods with persistent effect on future volatility distribution. The model statement is given by
\begin{eqnarray}\label{eq:m3.2}
 r_t &=& \mu_3 + K_{1t}J_{1t} + \exp\left(\frac{h_t}{2}\right) \varepsilon_t \\ \nonumber
 h_{t} &=& K_{2t}J_{2t} + \alpha + \phi(h_{t-1} - \alpha) + \sigma\eta_t,
\end{eqnarray}
where $(\varepsilon_t, \eta_t)$ follows a bivariate normal with mean zero, variance 1 and correlation $\rho$, the jump coefficients $K_{it}\sim N(\nu_i, \tau^2_i)$, the jump components $J_{it} \sim Ber(\pi_i)$ and $K_{it}, J_{it}$ are all independent with each other and with $\varepsilon_t$ and $\eta_t$. As in M2.3, we assume both $\nu_1=\nu_2=0$ and $\tau_1=\tau_2=1$, however, since $h_{t+1}$ is replaced by $h_{t}$ in the volatility modelling part, $h_t$ and $\varepsilon_t$ are now correlated (unlike M2.3), thus $\nu_1=0$ is not sufficient to facilitate EMH. Subsequently, the mean-correction term is given by 
$$ \mu_3 = -\left[ \frac{\rho\sigma}{2}\exp\left(\frac{\alpha}{2}+\frac{\sigma^2}{8(1-\phi^2)}\right) P\left(\frac{1}{2}\right) \right],$$
where 
$$ P(d)=\prod_{j=0}^\infty\left(1-\pi_2+\pi_2\exp\left\{ \frac{d^2\phi^{2j}}{2} \right\}\right).
$$

Table~\ref{tab:SVmoments} summarizes the first four central moments of the unconditional marginal distribution of $r_t$. Similar to M2.3, the marginal distributions of $h_t$ is no longer a normal, but, its mean is $\alpha$ and the variance is $(\sigma^2+\pi_2(1-\pi_2))/(1-\phi^2)$.

\section{Theoretical Comparison}\label{sec:theocomp}

This section presents a comparison of the six SVMs based on their abilities to correctly capture skewness, kurtosis and lead-lag correlations. These summary statistics measure crucial financial features like leverage, predictability, return-volatility asymmetric interaction and extreme observations.

Since all SVMs presented here satisfy EMH, the first unconditional moment of $r_t$ is zero, and the raw and central moments are same. Let $m_k^{(i.j)}$ denote the $k$-th order moment of $[r_t|\filtration]$ under $Mi.j$, for $i=2,3$ and $j=1,2,3$. Then the skewness and kurtosis are,
$$Sk_{(i,j)} = m_3^{(i.j)}\big/\left(m_2^{(i.j)}\right)^{(3/2)} \quad \textrm{and}\quad \kappa_{(i.j)} = m_4^{(i.j)}\big/\left(m_2^{(i.j)}\right)^2,$$
respectively. Table~\ref{tab:SVmoments} summarizes the theoretical expressions of three summary statistics (variance, skewness and kurtosis) for all six SVMs. 

Given the equivalence between the third and fourth moments, and skewness and kurtosis, we are not explicitly reporting the third and fourth order moments of $r_t$ distribution. 
The detailed derivation and proofs have not been reported due to length constraint. The expressions for skewness and kurtosis of M3.2 and M3.3 are left in terms of the respective lower order moments, and have not been explicitly worked out due to excessively long cumbersome terms. Nonetheless, the expressions presented in Table~\ref{tab:SVmoments} provide ready comparison between the two classes of SVMs. A few quick remarks are as follows:

\begin{table}[h!]\centering\caption{Variance, Skewness and Kurtosis of $[r_t|\filtration]$ under the six models $Mi.j$.}
\footnotesize 
\begin{tabular}{c|c|c} 
\hline
\hline 
 & M2.1 & M3.1 \\
 \hline
 \hline
 $m_2^{(i.1)}$ & $\exp\left\{\alpha + \frac{\sigma^2}{2(1-\phi^2)}\right\}$ & $\exp\left\{\alpha + \frac{\sigma^2}{2(1-\phi^2)}\right\} \times $ \\
& & $\left(1 + \rho^2 \sigma^2-\frac{\rho^2 \sigma^2}{4}\times\exp\left\{- \frac{\sigma^2}{4(1-\phi^2)}\right\}\right)$ \\ 
& & \\ 
 $Sk_{(i.1)}$ & 0 & $\frac{3}{2}\rho\sigma \exp\left\{\frac{3\alpha}{2}+\frac{9\sigma^2 }{8(1-\phi^2)}\right\} \left[-\frac{\rho^2\sigma^2}{3}\exp\left\{-\frac{3\sigma^2 }{4(1-\phi^2)}\right\} \right.$\\
& & $  \left. + 3+ \frac{9\sigma^2\rho^2}{4} -\left(1+\rho^2\sigma^2 \right) \exp\left\{-\frac{\sigma^2 }{2(1-\phi^2)}\right\}  \right] \big/ {(m_2^{(3.1)})^{3/2}}$\\ 
& & \\ 
 $\kappa_{(i.1)}$ & $3 \exp\left\{\frac{\sigma^2}{(1-\phi^2)}\right\}$ & $\exp\left\{2\alpha+\frac{2\sigma^2}{(1-\phi^2)} \right\} \left[3+16\rho^4\sigma^4 -9\rho^2\sigma^2\left(1+\frac{3\rho^2\sigma^2}{4} \right)\right. $ \\
& & $\times \exp\left\{ -\frac{3\sigma^2}{4(1-\phi^2)} \right\} -\frac{3\rho^4\sigma^4}{16}\exp\left\{ -\frac{3\sigma^2}{2(1-\phi^2)} \right\} +$ \\
& & $\left. 24\rho^2\sigma^2+\frac{3\rho^2\sigma^2}{2} (1+\rho^2\sigma^2)\exp\left\{ -\frac{5\sigma^2}{4(1-\phi^2)} \right\}\right] \big/ (m_2^{(3.1)})^2 $ \\
\hline 
\hline 
 & M2.2 & M3.2 \\
\hline 
\hline
 $m_2^{(i.2)}$ & $\exp\left\{\alpha + \frac{\sigma^2}{2(1-\phi^2)}\right\}  \omega^2 \xi_{\nu}(1)\left[1-\frac{2\delta^2}{\pi}\right]$&  $\exp\left\{\alpha + \frac{\sigma^2}{2(1-\phi^2)}\right\}  \omega^2 \xi_{\nu}(1) \times$ \\ 
 & & $\left[ \left( 1-\frac{2\delta^2}{\pi} \right)+\rho^2\sigma^2 (1-\delta^2) \right]-\mu_2^2$ \\ 
 & & \\ 
 $Sk_{(i,2)}$ & $\exp\left\{\frac{3\sigma^2}{8(1-\phi^2)}\right\}  \xi_{\nu}\left(\frac{3}{2}\right) \delta^3 \sqrt{\frac{2}{\pi}}\left[\frac{4}{\pi}-1\right]$ 
& $\left[3\mu_2 m_2^{3.2} +\mu_2^3+\omega^3 \xi_\nu(3/2) \exp\left( \frac{3\alpha}{2}+\frac{9\sigma^2}{8(1-\phi^2)} \right) \right.$ \\
& $\times \left(\xi_{\nu}(1) \left[1-\frac{2\delta^2}{\pi}\right]\right)^{-3/2} $  & $\times \left\{ \delta^3 \sqrt{\frac{2}{\pi}}\left( \frac{4}{\pi} -1\right) + \frac{9}{2}\rho\sigma\left( 1-\frac{2}{\pi} \right) \delta^2 \sqrt{1-\delta^2} \right. $\\ 
& & $\left.\left. +\frac{9}{2}\rho\sigma(1-\delta^2)^{3/2}\left( 1+\frac{3}{4}\rho^2\sigma^2\right)\right\}\right] \big/ \left(m_2^{(3.2)}\right)^{3/2} $ \\ 
& & \\ 
 $\kappa_{(i.2)}$ & $\exp\left\{\frac{\sigma^2}{(1-\phi^2)}\right\}  \left[3+\frac{8\delta^4}{\pi}-\frac{12\delta^4}{\pi^2}-\frac{12\delta^2}{\pi}\right]$ & $\left\{ 4\mu_2m_3^{3.2}-\mu_2^4-6\mu_2^2m_2^{3.2}+ \omega^4 \xi_\nu(2)\exp\left\{ 2\alpha+\frac{2\sigma^2}{1-\phi^2} \right\} \right.$ \\
  & $\times \xi_{\nu}(2) \left(\xi_{\nu}(1) \left[1 - \frac{2\delta^2}{\pi}\right]\right)^{-2}$ & $\times \left[ \delta^4 \left(3-\frac{4}{\pi}-\frac{12}{\pi^2}\right) +8\rho\sigma \delta^3 \sqrt{1-\delta^2}\sqrt{\frac{2}{\pi}}\left(\frac{4}{\pi}-1\right)\right. $ \\
	& & $\left. +6\delta^2(1-\delta^2)\left(1-\frac{2}{\pi} \right) (1+4\rho^2\sigma^2) +(1-\delta^2)^2\right. $ \\
	& & $ \left.\left. \times ( 3+6\rho^2-3\rho^4 +24\rho^2\sigma^2+16\rho^4\sigma^4)\right]\right\} \big/ \left(m_2^{(3.2)}\right)^2$ \\
\hline 
\hline 
 & M2.3  & M3.3 \\
\hline 
\hline
 $m_2^{(i.3)}$ & $\pi_1 + \exp\left\{\alpha + \frac{\sigma^2}{2(1-\phi^2)}\right\} P\left(1 \right)$ & $\pi_1+\exp\left\{\alpha+\frac{\sigma^2}{2(1-\phi^2)}\right\}\left( 1+\rho^2 \sigma^2 \right) P(1) -\mu_3^2$\\ 
		& &  \\ 
 $Sk_{(i.3)}$ & 0 & $\left[ 3m_2^{3.3}\mu_3+\mu_3^3+\frac{3\rho\sigma}{2}\pi_1 \exp\left\{ \frac{\alpha}{2}+\frac{\sigma^2}{8(1-\phi^2)} \right\}P\left( \frac{1}{2} \right)+ \right. $ \\
& & $ \left. \frac{9\rho\sigma}{2}\left( 1+\frac{3\rho^2\sigma^2}{4} \right)P\left( \frac{3}{2} \right)\exp\left\{ \frac{3\alpha}{2}+\frac{9\sigma^2}{8(1-\phi^2)} \right\}\right] \big/ (m_2^{(3.3)})^{3/2}$\\ 
		& &  \\ 
 $\kappa_{(i.3)}$ & $\left[ 3\pi_1+6\pi_1\exp\left\{ \alpha+\frac{\sigma^2}{2(1-\phi^2)} \right\}P(1)\right.$ & $\left[4\mu_3 m_3^{3.3}-\mu_3^4-6\mu_3^2 m_2^{3.3}+3\pi_1  +\exp\left\{\alpha+\frac{\sigma^2}{2(1-\phi^2)}  \right\}\right.$\\
& $\left. + 3\exp\left\{ 2\alpha+\frac{2\sigma^2}{1-\phi^2} \right\}P(2)\right]\big/(m_2^{(2.3)})^2$ & $\times 6\pi_1 P(1)(1+\rho^2\sigma^2)+\exp\left\{ 2\alpha +\frac{2\sigma^2}{1-\phi^2}\right\}P(2) $ \\
&  &$\left.\times(3+6\rho^2-3\rho^4 +24\rho^2\sigma^2+16\rho^4\sigma^4)\right] \big/ (m_2^{(3.3)})^2 $  \\
\hline 
\hline 

 \end{tabular} \label{tab:SVmoments}
\end{table}

\textbf{Remark 1.} \emph{The marginal return distribution under M2.1 and M2.3 are symmetric, irrespective of the parameter values including $\rho$.} This is perhaps undesirable as a symmetric return distribution would imply $Var(r_t|r_t>0) = Var(r_t|r_t<0)$, i.e., no leverage.

\textbf{Remark 2.} \emph{None of the summary statistics (variance, skewness and kurtosis) of the marginal return distribution under the $h_{t+1}$-based models (i.e., M2.1, M2.2 and M2.3) depend on $\rho$.} Though it is an interesting observation, it is not surprising, as $r_t$ depends on $h_t$ and $\varepsilon_t$, whereas, $h_t$ depends on $\eta_{t-1}$, and $\varepsilon_t$ and $\eta_{t-1}$ are uncorrelated.

Though the leverage effect was considered as the reason behind return skewness in the literature, Table~1 shows contradictory results for the $h_{t+1}$-based models. Moreover, the return skewness in model M2.2, is guided by $\delta =\lambda/\sqrt{1+\lambda^2}$, which is exogenous to return-volatility reaction mechanism. 

\textbf{Remark 3.} \emph{Assuming uncorrelated errors $(\varepsilon_t, \eta_t)$, none of the $h_t$-based SVMs (M3.1, M3.2, M3.3) requires mean-correction, that is, $\rho=0$ implies $\mu_1 = \mu_2 = \mu_3 = 0$. More importantly, substituting $\rho=0$ in the expressions for variance, skewness and kurtosis of $M3.j$ gives the corresponding expressions under $M2.j$.} In other words, under $h_{t+1}$-based models, there is no contribution of leverage in explaining extreme returns caused by anticipated jump in future volatility, whereas $h_t$-based models accounts for the contribution of leverage effect in generating extreme returns in presence of volatility jump.

\textbf{Remark 4.} \emph{While comparing the summary statistics for the two classes of SVMs, Remark 3 implies that  $\kappa(3.1) \ge \kappa(2.1)$.} For proving this result, it is sufficient to show that the terms of $\kappa(3.1)$ that contains $\rho$ is positive, i.e., 
\begin{eqnarray*}
&& 24\rho^2\sigma^2 + 16\rho^4\sigma^4 - 9\rho^2\sigma^2\left(1 + \frac{3\rho^2\sigma^2}{4}\right)\exp\left(-\frac{3\sigma^2}{4(1-\phi^2)}\right) \\
&& - \frac{3}{16}\rho^4\sigma^4\exp\left(-\frac{3\sigma^2}{2(1-\phi^2)}\right) + \frac{3}{2}\rho^2\sigma^2(1+\rho^2\sigma^2)\exp\left(-\frac{5\sigma^2}{4(1-\phi^2)}\right)\quad \ge 0.
\end{eqnarray*} 
This follows from the fact that $\exp(-x^2) \le 1$ for any real $x$. We believe that it may not be too difficult to show that $\kappa(3.j) \ge \kappa(2.j)$ for $j=2$ and $3$ as well. Since the expressions for $\kappa(3.2)$ and $\kappa(3.3)$ are complex functions of the respective lower order moments, it would require cumbersome calculations to segregate the terms that contain $\rho$.


We now use lead-lag correlations between $r_t$ and $e^{h_{t\pm k}}$ for the adequacy comparison of the two classes of SVMs. In spirit of \citeasnoun{Bollerslev2006}, we define model implied lead-lag correlations as $\rho_{\pm k}^{(i.j)} = Corr(r_t, e^{h_t\pm k})$ under model $Mi.j$, for $i=2,3$ and $j=1,2,3$. Then, 
$$ \rho_{\pm k}^{(i.j)} = \frac{Cov(r_t, e^{h_{t \pm k}}|\filtration)}{\sqrt{Var(r_t|\filtration)}\sqrt{Var(e^{h_t})}},$$
where the closed form expressions for $Var(r_t|\filtration)$ for different SVMs are given in Table~\ref{tab:SVmoments}.  Since the unconditional distribution of $h_t$ does not depend on $t$, the formula for $\rho_{\pm k}^{(i.j)}$ contains $Var(e^{h_t})$ and not $Var(e^{h_{t \pm k}})$. Moreover, the marginal distributions of $h_t$ for M2.1, M3.1, M2.2 and M3.2 are same, i.e., $N(\alpha, \sigma^2/(1-\phi^2))$, and in this case, 
\begin{equation}
 Var(e^{h_t})= \exp\left\{ 2\alpha+\frac{\sigma^2}{1-\phi^2}\right\}\left( \exp\left\{ \frac{\sigma^2}{1-\phi^2} \right\}-1 \right) ,
 \label{eq:var_eht_base}
\end{equation}
whereas, for M2.3 and M3.3, the marginal distributions of $h_t$ are same, but not normal. However, the long-range mean is still $\alpha$ and variance is $(\sigma^2 + \pi_2(1-\pi_2))/(1-\phi^2)$. One can also show that under these two jump models,  
\begin{equation}
 Var(e^{h_t})= \exp\left\{ 2\alpha+\frac{\sigma^2}{(1-\phi^2)}\right\}\left[ \exp\left\{ \alpha+\frac{\sigma^2}{1-\phi^2} \right\}P(2)-P^2(1) \right].
 \label{eq:var_eht_jump}
\end{equation}

As a result, deriving $Cov(r_t,e^{h_{t \pm k}}|\filtration)$ is sufficient for finding different lead-lag correlations. Furthermore, since $E(r_t|\filtration)=0$, the desired covariances simplify to $E[r_t e^{h_{t \pm k}}|\filtration]$ and here-onwards denoted as $\gamma_{\pm k}^{(i.j)}$ for $Mi.j$ and $k\ge 0$. Table~\ref{tab:Leadlag} summarizes the lead-lag covariance expressions (the proofs and derivations have been omitted due to length constraint). A few quick observations are as follows:

\textbf{Remark 5.} \emph{For all $h_{t+1}$-based SVMs presented in Section~2, the contemporaneous and lagged covariances and hence correlations are zero, i.e., $\rho_{-k}^{(i.j)}=0$ for $k\ge 0$.} Trivially, one can state that the $h_t$-based SVM gives larger contemporaneous correlation than the corresponding $h_{t+1}$-based SVM, that is, $\rho_0^{(3.j)} > \rho_0^{(2.j)}$ for each $j=1,2,3$. A closer look at the table shows that all lead and lagged-covariances and hence the corresponding correlations asymptote to zero as $k \rightarrow\infty$.

\textbf{Remark 6.} For $j=1,2,3$, (a) $\gamma_{+k}^{(2.j)}$ are multiples of $\phi^{k-1}$, (b) $\gamma_{\pm k}^{(3.j)}$ have very specific pattern, and a common term involving  $\exp(-\sigma^2\phi^k/(2(1-\phi^2)))$, for all $k$, is being adjusted due to the additional mean-correction term.

\textbf{Remark 7.} For a fixed non-zero $k$, though the comparison between $\gamma_{\pm k}^{2.j}$ and $\gamma_{\pm k}^{3.j}$ can be made with some effort, the comparison between $\rho_{\pm k}^{2.j}$ and $\rho_{\pm k}^{3.j}$ requires conditions on model parameters as the marginal variances of $h_t$ and $r_t$ may vary with models.

\begin{table}[H]\centering \caption{List of lead-lag covariances $\gamma_{\pm k}^{(i.j)}$, for $i=2,3$ and $j=1,2,3$, for the two classes of SVMs. Here, $\rho \sigma \exp\left\{ \frac{3\alpha}{2} + \frac{\sigma^2(5 + 4\phi^k)}{(8(1-\phi^2))}\right\}$ is a common factor in each expression.}
\begin{tabular}{c|c|c}
\hline
 & M2.1 & M3.1 \\
 \hline
$\gamma_0^{(i.1)}$ & 0 & $\frac{3}{2} - \frac{1}{2}\exp\left(\frac{-\sigma^2}{2(1-\phi^2)}\right)$\\
& & \\
$\gamma_{+k}^{(i.1)}$ & $\phi^{k-1}$ & $\left(\phi^k+\frac{1}{2}\right) - \frac{1}{2}\exp\left(\frac{-\sigma^2\phi^k}{2(1-\phi^2)}\right)$\\
& & \\
$\gamma_{-k}^{(i.1)}$ & 0 & $\frac{1}{2} - \frac{1}{2}\exp\left(\frac{-\sigma^2\phi^k}{2(1-\phi^2)}\right) $\\
\hline
\hline
 & M2.2 & M3.2 \\
 \hline
$\gamma_0^{(i.2)}$ & 0 & $ \left[\omega\sqrt{1-\delta^2}\xi_{\nu}(\frac{1}{2})\right] \left(\frac{3}{2} - \frac{1}{2}\exp\left(\frac{-\sigma^2}{2(1-\phi^2)}\right)\right) $\\
& & \\
$\gamma_{+k}^{(i.2)}$ & $\phi^{k-1} \left[\omega\sqrt{1-\delta^2}\xi_{\nu}(\frac{1}{2})\right]$ & $ \left[\omega\sqrt{1-\delta^2}\xi_{\nu}(\frac{1}{2})\right]  \left(\left(\phi^k+\frac{1}{2}\right)  - \frac{1}{2}\exp\left(\frac{-\sigma^2\phi^k}{2(1-\phi^2)}\right)\right)$\\
& & \\
$\gamma_{-k}^{(i.2)}$ & 0 & $ \left[\omega\sqrt{1-\delta^2}\xi_{\nu}(\frac{1}{2})\right]  \left(\frac{1}{2} - \frac{1}{2}\exp\left(\frac{-\sigma^2\phi^k}{2(1-\phi^2)}\right)\right)$\\
\hline
\hline
 & M2.3  \tablefootnote{$P_{k-1}(d)=\prod_{j=0}^{k-1}\left(1-\pi_2+\pi_2\exp\left\{ \frac{d^2\phi^{2j}}{2} \right\}\right)$} & M3.3 \\
 \hline
$\gamma_0^{(i.3)}$ & 0 & $\frac{3}{2}P\left(\frac{3}{2}\right) -  P(1)P\left(\frac{1}{2}\right)\frac{1}{2}\exp\left(\frac{-\sigma^2}{2(1-\phi^2)}\right) $\\
 & & \\
$\gamma_{+k}^{(i.3)}$ & $\phi^{k-1} P\left(\phi^k + \frac{1}{2}\right) P_{k-1}(1)$  & $\left(\phi^k+\frac{1}{2}\right)  P\left(\phi^k + \frac{1}{2}\right) P_{k-1}\left(1\right) - P(1)P\left(\frac{1}{2}\right)\frac{1}{2}\exp\left(\frac{-\sigma^2\phi^k}{2(1-\phi^2)}\right)$ \\
 & & \\
$\gamma_{-k}^{(i.3)}$ & 0 & $\frac{1}{2}  P\left(\frac{\phi^k}{2} + 1\right) P_{k-1}\left(\frac{1}{2}\right) - P(1)P\left(\frac{1}{2}\right)\frac{1}{2}\exp\left(\frac{-\sigma^2\phi^k}{2(1-\phi^2)}\right)$ \\
\hline
\end{tabular}\label{tab:Leadlag}
\end{table}


\section{Application to real-life data \& Comparison} \label{sec:app}

In this section we compare the implementation performance of the two classes of SVMs on four real-life data sets, (a) the oil reference basket (ORB) price provided by the Organization of the Petroleum Exporting Countries (OPEC), (b) stock price of CITI bank, (c) British Euro vs. United States Dollar exchange rate, and (d) S\&P 500 index. These data were selected in such a way that empirical contemporaneous correlations take both positive value (S\&P 500) and negative value (for the other three data sets).

Following \citeasnoun{Jacquier2004185}, \citeasnoun{Berg2004107} and \citeasnoun{Liu2014}, we use a Bayesian framework for estimating parameters involved in these SVMs. As in \citeasnoun{Nakajima20092335}, our Markov Chain Monte Carlo (MCMC) algorithms used the following priors for the common parameters ($\alpha, \phi, \sigma$ and $\rho$) in the six models:
\begin{eqnarray*}
	\alpha &\sim&  N(0,1), \hspace{0.7cm} \frac{\phi+1}{2}\sim Beta(20,1.5), \\ \frac{1}{\sigma^{2}}&\sim& \Gamma(2.5,0.025), \hspace{0.5cm} \rho\sim U(-1,1). 
\end{eqnarray*}
For the skewed-$t$ models (M2.2 and M3.2), the degrees of freedom parameter, $\nu$, has an exponential prior with mean hyperparameter $10$ which supports a heavy-tailed return distribution, and the skewness parameter $\lambda$ has a fairly non-informative prior (i.e., normal with mean $0$ and unit variance). For the jump models (M2.3 and M3.3) we observe that the jumps are rare, and hence we assume $Beta(2,100)$ prior for both jump probabilities $\pi_1$ and $\pi_2$.
 
All model implementation codes were run on a 4-core 3.7GHz Accelerated Processing Unit (APU) using Just Another Gibbs Sampler (JAGS -3.4.0) with parallel processing from within R. The plug-in estimates of the parameters were obtained by first throwing away a burn-in of 10,000 initial posterior realizations, and then let the MCMC run until the chains converged, that is, the multivariate \emph{potential scale reduction factor} (psrf) value is close to 1 \cite{Gelman1992}. The final length of the chain is calculated using Raftery and Lewis's criteria \cite{raftery1992}.  For the log-volatility estimates $(\hat{h}_t)$, used in computing lead-lag correlations, we used 30,000 posterior realizations for each of the time points of returns. 

We measure the performance comparison in terms of persistence in time-varying volatility, lead-lag correlation between return and volatility, skewness and heavy tail of return distribution. Following the findings of \citeasnoun{Bollerslev2006}, \citeasnoun{ait2013} and \citeasnoun{wang2014}, we compare the models with respect to the return-volatility correlation. In an empirical manner, we assume $r_t^2$ as a proxy of the unobservable volatility, and approximate lead-lag correlations by $Corr(r_t, r_{t\pm k}^2)$ (referred to as \emph{empirical correlation}). Alternatively, we use the MCMC realizations of $h_t$ to estimate $Corr(r_t, e^{h_t})$ (called as \emph{model estimated correlations}). 

\subsection{ORB Price data}

Oil price plays an important role as a macro-economic indicator. For example, large increase in oil price (or positive return) leads to the rise in the production cost and hence the reduction in the GDP \cite{Rotemberg1999}. Similar to the stock based financial derivatives, volatility of oil return also plays a crucial role in determining prices of oil derivatives.

\begin{figure}[h!]\centering
\subfigure[Return series]{\includegraphics[width=2.75in]{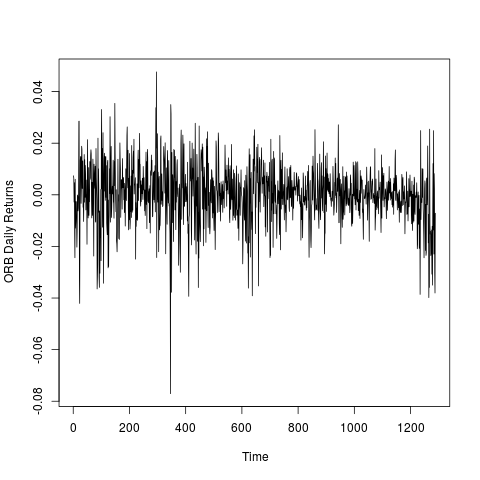}}
\subfigure[Density plot]{\includegraphics[width=2.75in]{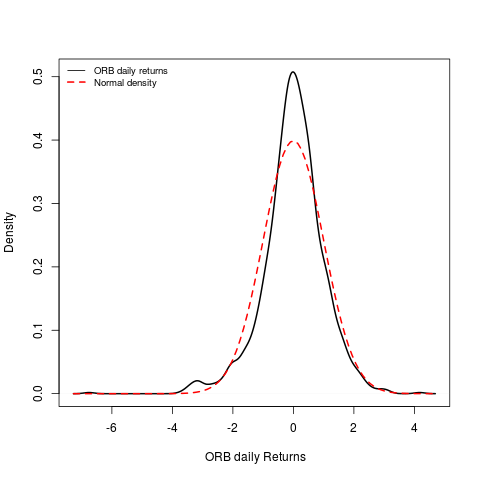}}\vspace{-0.1in}
\subfigure[Empirical lead-lag correlation]{\includegraphics[width=2.75in]{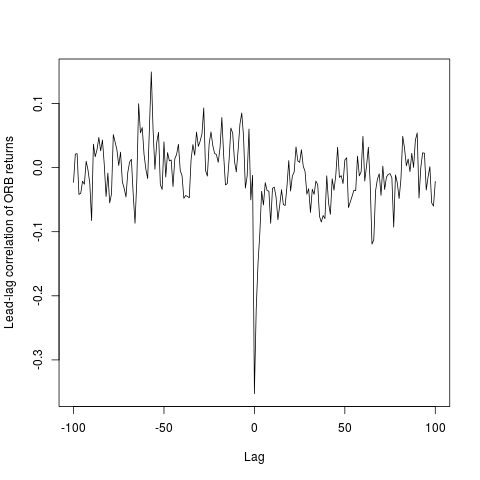}}
\subfigure[Normal QQ-plot]{\includegraphics[width=2.75in]{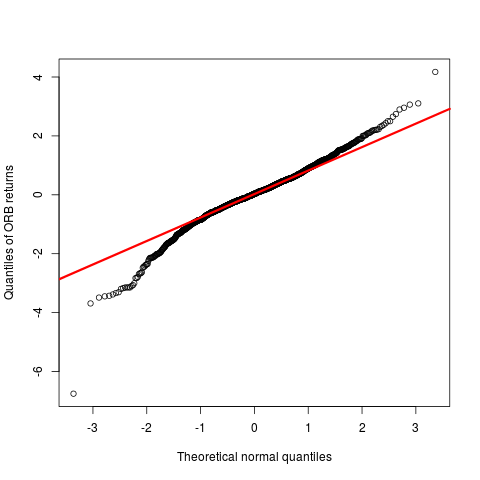}}\caption{Exploratory plots of daily ORB return price between January 1, 2010 and January 1, 2016 obtained from OPEC.}\label{fig:Oil_summary}
\end{figure}

In this example we apply the six models discussed above on the ORB price obtained from OPEC. The ORB price is a weighted average of a basket of oil prices obtained from different oil producing countries. The correlation between $r_t$ and $h_t$ can be justified as the reduction in the oil price increases the risk of reducing the revenue of oil-exporting countries contributing in the reference basket. We analyse the returns, difference of logarithms of 1289 daily ORB prices, between January 1, 2010 to January 1, 2016. Figure~\ref{fig:Oil_summary} presents the empirical summaries.

The return data has mean $-0.3260 \times 10^{-3}$, which was adjusted to obtain the mean zero series. Other defining sample moments are standard deviation $= 0.1141 \times 10^{-1}$, skewness ($ Sk = -0.5659$) and kurtosis ($\kappa = 5.7315$) of the return series. From Figure~\ref{fig:Oil_summary} and the summary statistics, it is clear that the data comes from a heavy-tail distribution, and the empirical lead-lag correlation, $Corr(r_t,r_t^2)$, is minimized at $k=0$. The empirical contemporaneous correlation is -0.3526. Simple Bartlett's confidence interval (-0.05635,0.05635) indicates that the contemporaneous correlation is significant. 

\begin{figure}[h!]\centering
    \includegraphics[width=3in]{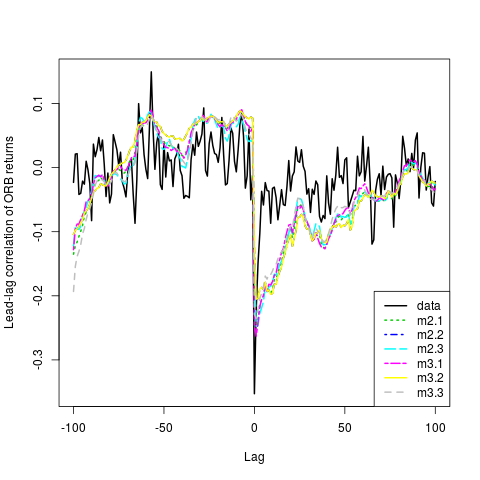}
	\includegraphics[width=3in]{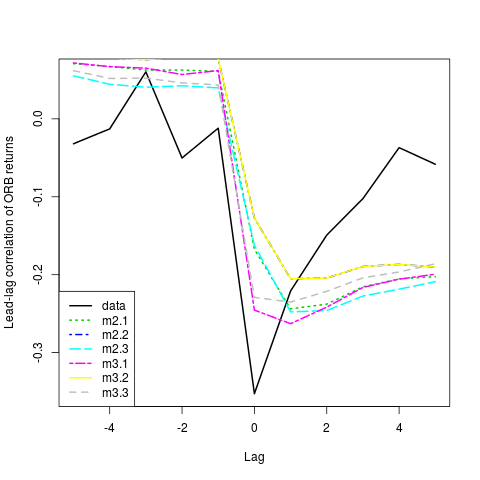}
     \caption{ORB: Comparison of model estimated lead-lag correlations, $Corr(r_t,e^{h_{t\pm k}}|\Theta, {\bf r})$ using MCMC realizations of $h_{t\pm k}$, and the empirical values, $Corr(r_t, r_t^2)$, for all six SVMs.}\label{fig:oil_post_leadlag}
\end{figure}

We investigated the results further and compared the \emph{model estimated lead-lag correlations} with the \emph{empirical lead-lag correlations} (see Figure~\ref{fig:oil_post_leadlag}).

According to Figure~\ref{fig:oil_post_leadlag}, all SVMs may appear to be doing a comparable job in capturing the overall nature. However, a closer investigation reveals that M3.1 and M3.3 give better approximation of the empirical values.

\begin{figure}[h!]\centering   
	\subfigure[$\alpha$]{\includegraphics[width=3.0in]{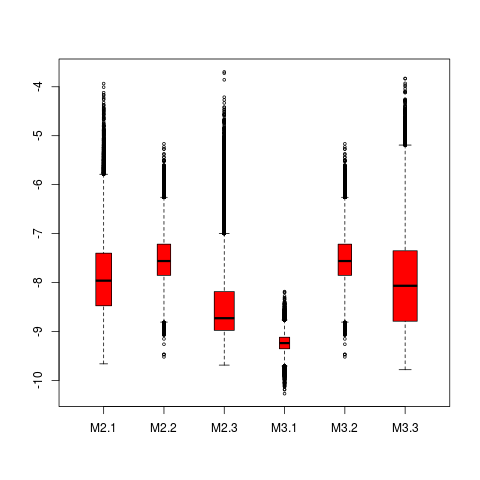}}
	\subfigure[$\sigma$]{\includegraphics[width=3.0in]{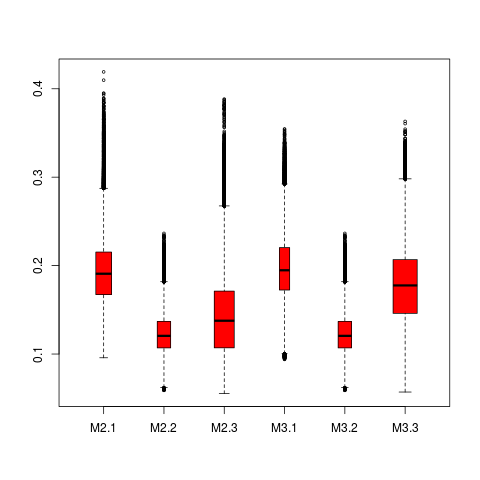}}\vspace{-0.1in}
	\subfigure[$\log(\phi/(1-\phi))$]{\includegraphics[width=3.0in]{Oil_sigma_all_models.png}}
	\subfigure[$\rho$]{\includegraphics[width=3.0in]{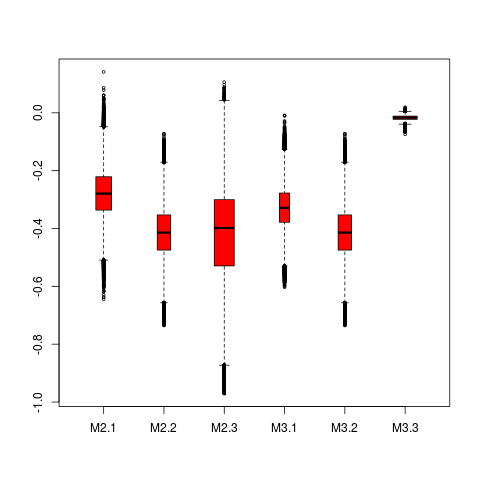}} 
\caption{Oil price data: posterior distribution of $\alpha,\sigma,\phi,\rho$ using the MCMC realizations obtained from JAGS for all six models in Sections 2 and 3.} \label{fig:oil_post_density_box}
\end{figure}

The parameter estimates (posterior mean and 95\% credible interval) of the six SVMs fitted to the ORB returns, implemented via JAGS, are reported in Table~\ref{tab:ORB_param}. For more thorough comparison of the common parameters ($\alpha, \phi, \sigma, \rho$) Figure~\ref{fig:oil_post_density_box} presents the posterior densities via box plots. 
\begin{table}[h!]\centering	
\caption{Bayesian estimates of model parameters for both $h_t$-and $h_{t+1}$-based SVMs}
\begin{scriptsize}	
		\begin{tabular}{|c|c|c|c||c|c|c|}
			\hline Parameters & M2.1 & M2.2 & M2.3 & M3.1 & M3.2 & M3.3 \\ \hline
			$\alpha$   & -7.88  & -7.53 & -8.5 & -9.23 & -8.087 & -7.9936 \\ 
								 & (-9.2,-6.45) & (-8.389,-6.568) & (-9.37,-7.06)  & (-9.61,-8.86) & (-8.93,-7.14) & (-9.37,-6.37) \\ \hline
			$\sigma$ & 0.19 & 0.1228 & 0.14 & 0.2 & 0.15 & 0.18 \\
								 & (0.13,0.26) & (0.08,0.17) & (0.066,0.22) & (0.13,0.27) & (0.10,0.20) & (0.09,0.26) \\ \hline
			$\phi$ & 0.985 & 0.982 & 0.977 & 0.97 & 0.987 & 0.982 \\
								 & (0.969,0.99) & (0.96,0.99) & (0.95,0.99) & (0.95,0.9897) & (0.98,0.99) & (0.96,0.99) \\ \hline
			$\rho$ & -0.28 & -0.41 & -0.42 & -0.28  & -0.41 & -0.017 \\
							& (-0.43,-0.11) & (-0.59,-0.24) & (-0.80,-0.14) & (-0.48,-0.18) & (-0.58,-0.22) & (-0.034,-0.0032) \\ \hline
			$\nu$ & -- & 14.83 & -- & --  & 15.43 & -- \\
								 &  & (6.68,26.26) &  &   & (7.17,26.68) &  \\ \hline
			$\lambda$ & -- & -0.005 & -- & -- & 0.012 & -- \\
								 &   &  (-0.01,-0.0013) &  &  &  (-0.029,0.056) &  \\ \hline
			$\pi_1$ & -- & -- & 0.0015 & --  & -- & 0.0011 \\
								 &  & & (0.00002,0.036) &  & & (0.00005,0.0036) \\ \hline
			$\pi_2$ & -- & -- & 0.023 & --  & -- & 0.0187 \\
								 &  & & (0.0012,0.049) &  & & (0.0004,0.043) \\ \hline
		\end{tabular}
\end{scriptsize}\label{tab:ORB_param} 
\end{table}

Table~\ref{tab:ORB_param} and Figure~\ref{fig:oil_post_density_box} show that most of the estimated model parameters are comparable (i.e., statistically significantly indistinguishable) across the two types of ($h_t$ - vs. $h_{t+1}$-based) SVMs and in agreement with the results reported in other literature (see e.g., \citeasnoun{vo2009regime}).  A few interesting findings from Table~\ref{tab:ORB_param} and Figure~\ref{fig:oil_post_density_box} are as follows:

The long term volatility ($\alpha$) is estimated to be lower in magnitude in $h_{t+1}$-based model M2.1 as compared to the $h_t$-based model M3.1. MCMC estimates of the volatility of log-volatility process (measured by $\sigma$) exhibit an interesting pattern, i.e.,  $\sigma_{(i.1)} > \sigma_{(i.3)} > \sigma_{(i.2)}$, and $\sigma_{(2.j)} < \sigma_{(3.j)}$ for $i=2,3$ and $j=1,2,3$. The results show that high volatility clustering with estimated $\phi$ close to 1 is common to all the models and significant estimate of error correlation $\rho$ close to -0.3 is present for all models except for M3.3. It appears that the jump component in the log-volatility process of M3.3 overshadows the effect of correlation between $\varepsilon_t$ and $\eta_t$. This observation is consistent with other data sets as well (see Figures~\ref{fig:CITI_post_density_box}, \ref{fig:EURO_post_density_box}, \ref{fig:SNP_post_density_box}).


\subsection{CITI Price data}

\begin{figure}[h!]\centering
  \subfigure[Return series]{\includegraphics[width=2.65in]{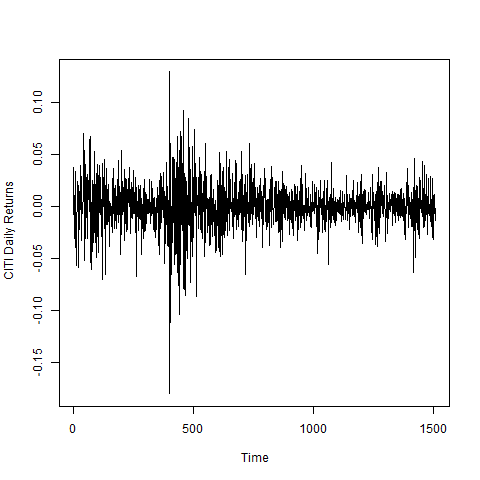}}
\subfigure[Density plot]{\includegraphics[width=2.65in]{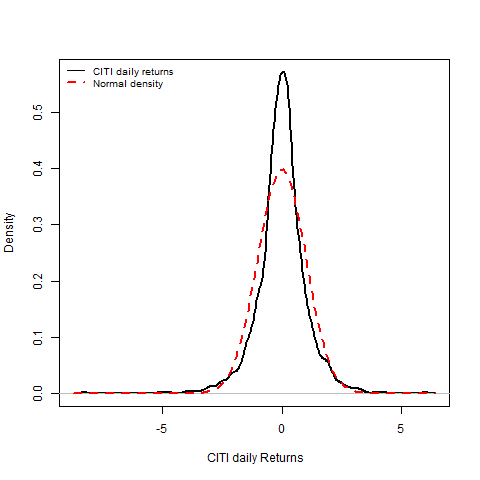}}
\subfigure[Empirical lead-lag correlation]{\includegraphics[width=2.65in]{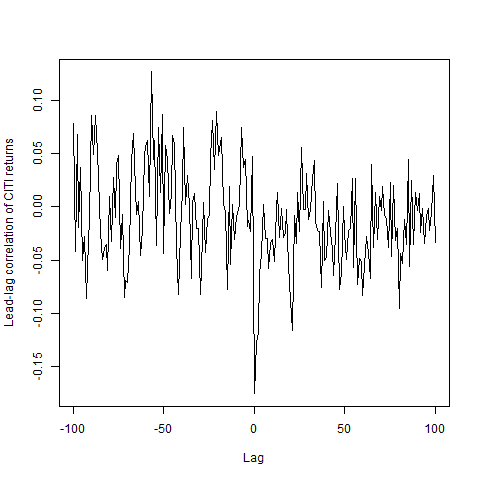}}
\subfigure[Normal QQ-plot]{\includegraphics[width=2.65in]{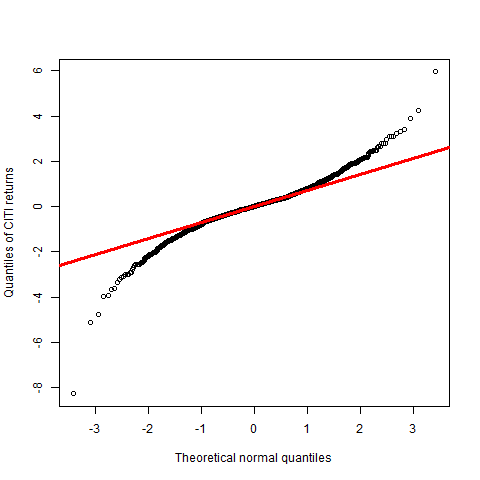}}
\caption{Exploratory plots of daily CITI return between Jan. 01, 2010 to Jan. 01, 2016.}\label{fig:CITI_summary}
\end{figure}

The daily stock prices of CITI bank (here-onwards referred to as CITI) obtained from New York Stock Exchange (NYSE) are modelled using the six SVMs considered here. The CITI returns $r_t = \log(P_t)-\log(P_{t-1})$, where $P_t$ denote 1509 daily prices between Januray 1, 2010 to January 1, 2016, are summarized in Figure~\ref{fig:CITI_summary}.

Important summary statistics of the mean-adjusted return series are: standard deviation, $0.218 \times 10^{-1}$, skewness, $ Sk = -0.4365$, and kurtosis, $\kappa = 8.7734$. That is, the return distribution seems to be negatively skewed and have sharp peak along with heavy tails. The empirical lead-lag correlation plot, Figure~\ref{fig:CITI_summary}(c), shows $\rho_0=-0.1755$ is maximum in magnitude, which is outside Bartlett's 95\% confidence interval about zero and hence significant.

\begin{figure}[H]\centering
    \includegraphics[width=2.75in]{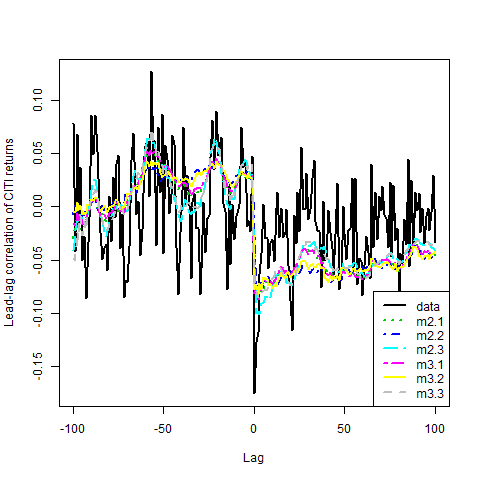}
	\includegraphics[width=2.75in]{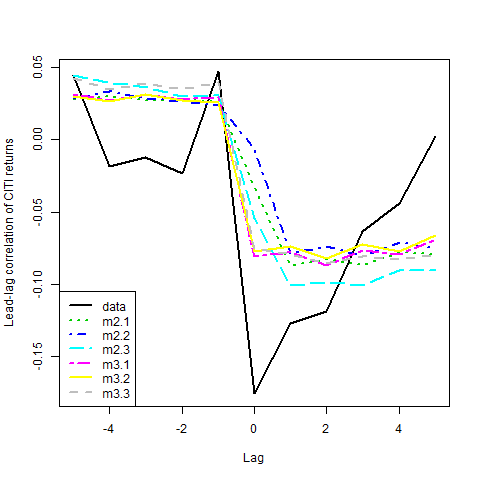}
     \caption{CITI: Comparison of model estimated lead-lag correlations, $Corr(r_t,e^{\hat{h}_{t\pm k}}|\Theta, {\bf r})$ using MCMC estimates of $h_{t\pm k}$, and the empirical values, $Corr(r_t, r_{t \pm k}^2)$, for all six SVMs.}\label{fig:CITI_post_leadlag}
\end{figure}

Figure~\ref{fig:CITI_post_leadlag} shows that the model estimated lead-lag correlation for $h_{t+1}$-based models provides closer approximation to the empirical lead-lag correlations as compared to the $h_t$ based SVMs. The exponential increase in estimated model implied lead correlations are evident from the second panel of this figure.

\begin{figure}[h!]\centering   
	\subfigure[$\alpha$]{\includegraphics[scale=0.45]{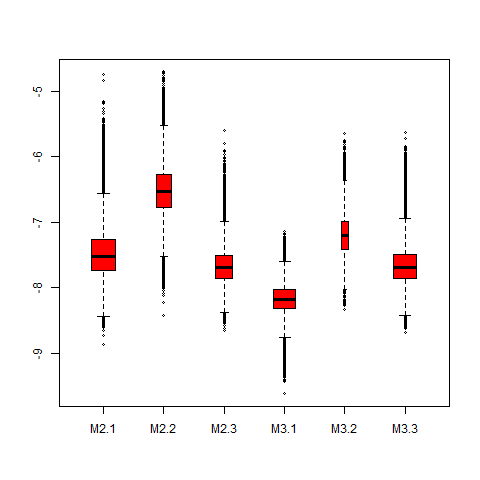}}
	\subfigure[$\sigma$]{\includegraphics[scale=0.45]{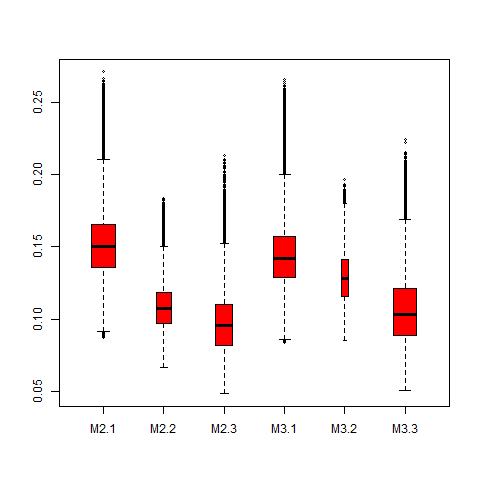}} 	
	\subfigure[$\log(\phi/(1-\phi))$]{\includegraphics[scale=0.45]{CITI_sigma_all_models.png}}
	\subfigure[$\rho$]{\includegraphics[scale=0.45]{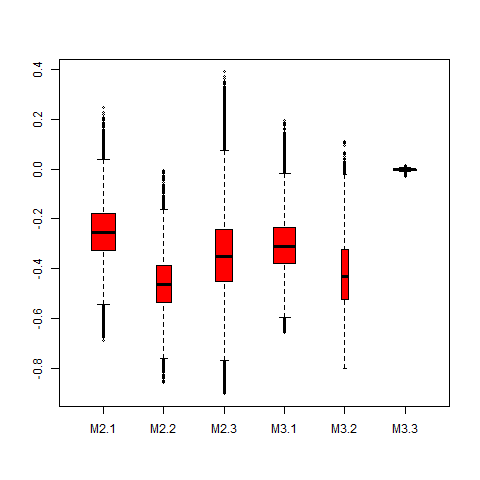}} 
\caption{CITI data: posterior distribution of $\alpha,\sigma,\phi,\rho$ based on the MCMC realizations for all models in Sections 2 and 4.} \label{fig:CITI_post_density_box}
\end{figure}

Table~\ref{tab:CITI_param} provides parameter estimates in terms of posterior means and 95\% credible intervals, and Figure~\ref{fig:CITI_post_density_box} depicts the posterior distributions of the common parameters. 

\begin{table}[h!]\centering	
\caption{Bayesian estimates of model parameters for both $h_t$-and $h_{t+1}$-based SVMs}
\begin{scriptsize}	
		\begin{tabular}{|c|c|c|c||c|c|c|}
			\hline Parameters & M2.1 & M2.2 & M2.3 & M3.1 & M3.2 & M3.3 \\ \hline
			$\alpha$   & -7.47  & -6.5 & -7.66 & -8.18 & -7.19 & -7.66 \\ 
								 & (-8.108,-6.75) & (-7.25,-5.77) & (-8.16,-7.1)  & (-8.604,-7.72) & (-7.79,-6.56) & (-8.17,-7.05) \\ \hline
			$\sigma$ & 0.152 & 0.1081 & 0.098 & 0.144 & 0.13 & 0.11 \\
								 & (0.1116,0.1988) & (0.079,0.138) & (0.006,0.14) & (0.1056,0.189) & (0.095,0.16) & (0.066,0.1539) \\ \hline
			$\phi$ & 0.9858 & 0.9853 & 0.985 & 0.9845 & 0.9878 & 0.9843 \\
								 & (0.9776,0.99) & (0.9769,0.99) & (0.9768,0.99) & (0.9756,0.99) & (0.9833,0.99) & (0.97,0.9899) \\ \hline
			$\rho$ & -0.25 & -0.4655 & -0.343 & -0.31  & -0.4181 & -0.0031 \\
							& (-0.46,-0.043) & (-0.6749,-0.2589) & (-0.66,-0.02) & (-0.51,-0.09) & (-0.676,-0.122) & (-0.0095,0.0029) \\ \hline
			$\nu$ & -- & 8.95 & -- & --  & 8.98 & -- \\
								 &  & (5.33,13.39) &  &   & (5.61,14.3) &  \\ \hline
			$\lambda$ & -- & -0.0081 & -- & -- & 0.0016 & -- \\
								 &   &  (-0.014,-0.0031) &  &  &  (-0.05,0.05) &  \\ \hline
			$\pi_1$ & -- & -- & 0.0015 & --  & -- & 0.0014 \\
								 &  & & (0.00003,0.0035) &  & & (0.00005,0.0033) \\ \hline
			$\pi_2$ & -- & -- & 0.016 & --  & -- & 0.0149 \\
								 &  & & (0.0023,0.03) &  & & (0.0024,0.031) \\ \hline
		\end{tabular}
\end{scriptsize}\label{tab:CITI_param} 
\end{table}

The parameter estimates and corresponding credible intervals are in concordance with the literature. We find similar patterns, as in ORB price data analysis results. Figure~\ref{fig:CITI_post_density_box} also shows that both $h_t$ and $h_{t+1}$-based SVMs result in statistically indistinguishable posterior distributions for most of the key parameters, except $\rho$ in M3.3 (similar to ORB data).

\subsection{EURO-US Dollar exchange rate}

\begin{figure}[h!]\centering
  \subfigure[Return series]{\includegraphics[width=2.95in]{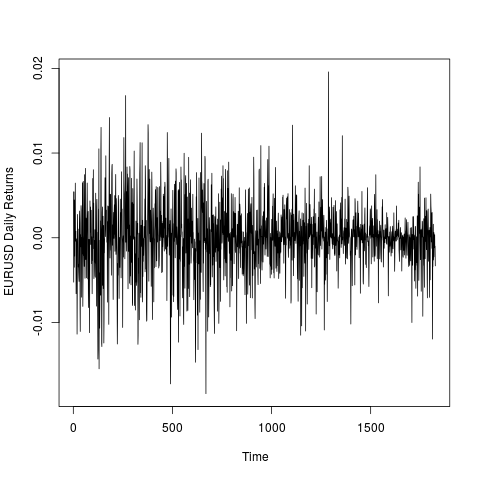}}
\subfigure[Density plot]{\includegraphics[width=2.95in]{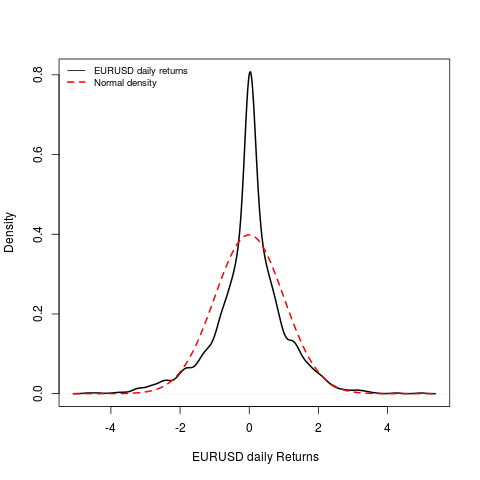}}
\subfigure[Empirical lead-lag correlation]{\includegraphics[width=2.95in]{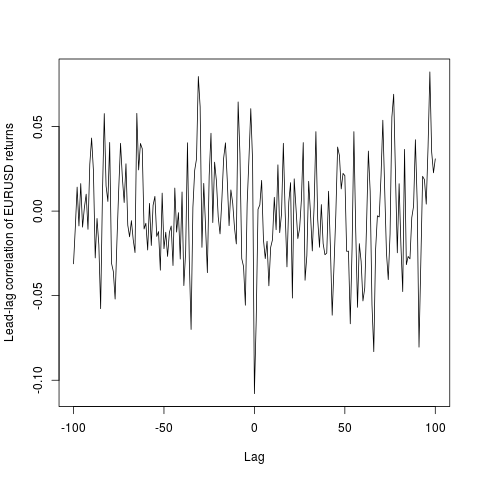}}
\subfigure[Normal QQ-plot]{\includegraphics[width=2.95in]{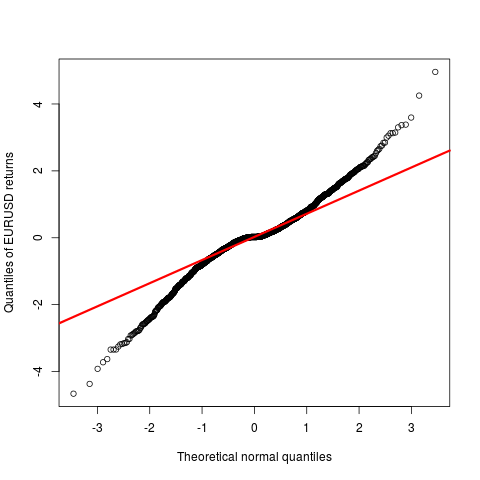}}
        \caption{Exploratory plots of daily EURO-USD return between 01/01/2010 --- 31/12/2014.}\label{fig:EURO_summary}
\end{figure}

Exchange rate is one of the major determinants of investments in a country. It is well known that the {foreign exchange market is much larger to equity markets, and thus risk} associated with it is of critical importance. Early literature in exchange rate volatility modeling describes that increase of one currency value is equivalent to decrease of the other in equal magnitude. This two sided nature of foreign exchange supports the symmetric return-volatility relationship, which further justifies the assumption of symmetric distribution for exchange rate return \cite{bollerslev1992}. However, recent literature shows presence of skewness in the exchange rate return distribution data. \citeasnoun{diebold1989} and more recently \citeasnoun{patton2006} and \citeasnoun{beine2007} have reported both positive and negative skewness in different exchange rate returns observed over different time intervals. In this section we apply all six SVMs on the 1826 returns on Euro to US Dollar exchange rates observed during January 01, 2010 and December 31, 2014. Figure~\ref{fig:EURO_summary} summarizes the interesting features of the data.

Important summary statistics of the mean-adjusted return series are: standard deviation, $0.0394 \times 10^{-1}$, skewness $ (Sk) = -0.2404$, and kurtosis $ = 5.1829$. That is, the return distribution is slightly negatively skewed and leptokurtic. The empirical lead-lag correlation plot, Figure~\ref{fig:EURO_summary}(c), shows largest value (magnitude) at lag zero, viz. $\rho_0=-0.108$ (as earlier, it is significant according to Bartlett's interval).

\begin{figure}[h!]\centering
    \includegraphics[width=2.85in]{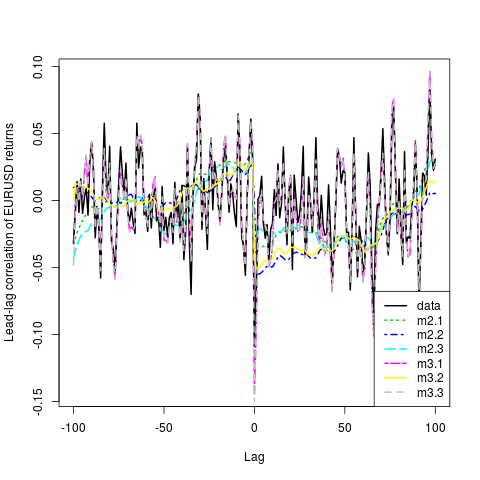}
	\includegraphics[width=2.85in]{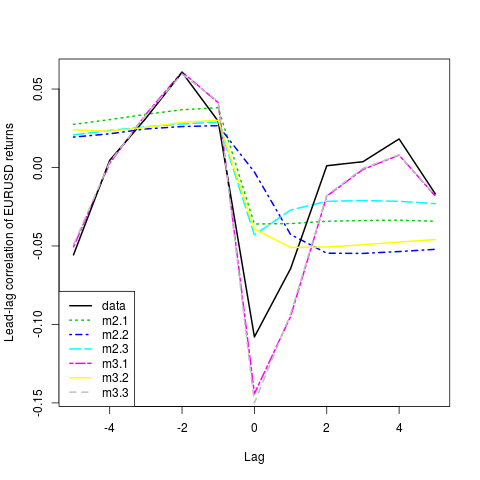}
     \caption{EURO-USD: Comparison of model estimated correlations, $Corr(r_t,e^{h_{t\pm k}}|\Theta, {\bf r})$ using MCMC realizations of $h_{t\pm k}$, and the empirical values, $Corr(r_t, r_t^2)$, for all six SVMs.}\label{fig:EURO_post_leadlag}
\end{figure}

Figure~\ref{fig:EURO_post_leadlag} depicts a thorough comparison of the lead-lag correlations (both model estimated and empirical) for the two classes of SVMs.

As in the previous examples, lead-lag correlation plots (Figure~\ref{fig:EURO_post_leadlag}) show that M3.1 and M3.3 give good approximations (in fact very good in this example) of the empirical values as compared to other models.

\begin{figure}[h!]\centering   
	\subfigure[$\alpha$]{\includegraphics[width=2.5in]{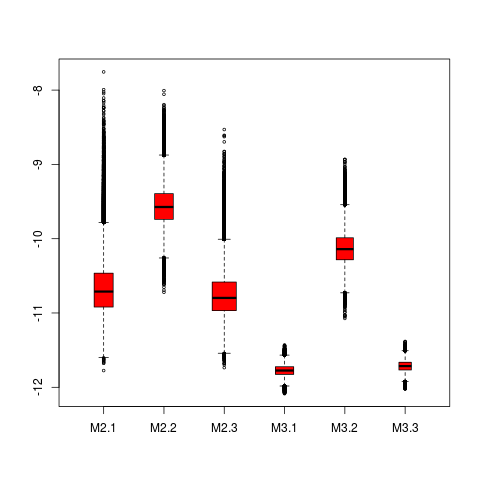}}
	\subfigure[$\sigma$]{\includegraphics[width=2.5in]{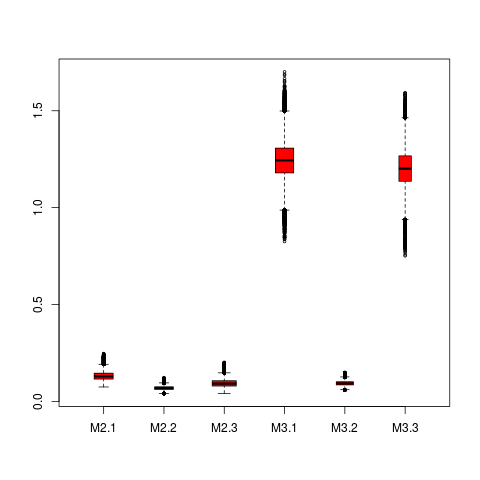}} 	
	\subfigure[$\log(\phi/(1-\phi))$]{\includegraphics[width=2.5in]{EURUSD_sigma_all_models.png}}
	\subfigure[$\rho$]{\includegraphics[width=2.5in]{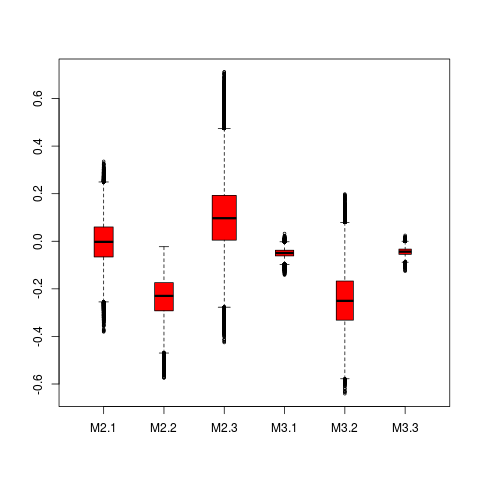}} 
\caption{EURO-USD data: posterior distribution of $\alpha,\sigma,\phi,\rho$ based on the MCMC realizations for all models in Sections 2 and 4.} \label{fig:EURO_post_density_box}
\end{figure}

Table~\ref{tab:EURO_param} summarizes the posterior means and 95\% credible intervals for the parameter of the two classes of SVMs, and the box plots in figure \ref{fig:EURO_post_density_box} describe the full posterior distribution of the common parameters, viz. $\alpha,\;  \phi, \; \sigma,\; \rho$. 
\begin{table}[h!]\centering	
\caption{Bayesian estimates of model parameters for both $h_t$-and $h_{t+1}$-based SVMs}
\begin{scriptsize}	
		\begin{tabular}{|c|c|c|c||c|c|c|}
			\hline Parameters & M2.1 & M2.2 & M2.3 & M3.1 & M3.2 & M3.3 \\ \hline
			$\alpha$   & -10.67  & -9.56 & -10.75 & -11.77 & -10.13 & -11.87 \\ 
								 & (-11.31,-9.94) & (-10.08,-9.01) & (-11.29,-10.13)  & (-11.93,-11.63) & (-10.56,-9.69) & (-11.87,-11.56) \\ \hline
			$\sigma$ & 0.13 & 0.068 & 0.094 & 1.24 & 0.09 & 1.2 \\
								 & (0.088,0.181) & (0.05,0.088) & (0.006,0.137) & (1.06,1.43) & (0.07,0.12) & (1.01,1.4) \\ \hline
			$\phi$ & 0.9873 & 0.9859 & 0.9855 & 0.44 & 0.989 & 0.46 \\
								 & (0.9814,0.99) & (0.978,0.99) & (0.976,0.99) & (0.3277,0.5535) & (0.9869,0.99) & (0.34,0.57) \\ \hline
			$\rho$ & -0.0037 & -0.24 & 0.1027 & -0.05  & -0.2473 & -0.044 \\
							& (-0.18,0.18) & (-0.4,-0.084) & (-0.178,0.3965) & (-0.089,-0.015) & (-0.48,-0.011) & (-0.0677,-0.013) \\ \hline
			$\nu$ & -- & 4.34 & -- & --  & 4.16 & -- \\
								 &  & (4,4.92) &  &   & (4,4.48) &  \\ \hline
			$\lambda$ & -- & -0.0064 & -- & -- & 0.015 & -- \\
								 &   &  (-0.012,-0.002) &  &  &  (-0.042,0.071) &  \\ \hline
			$\pi_1$ & -- & -- & 0.0016 & --  & -- & 0.00105 \\
								 &  & & ($1.2\times 10^{-4}$,0.0034) &  & & ($2.75\times 10^{-5}$,0.0025) \\ \hline
			$\pi_2$ & -- & -- & 0.015 & --  & -- & 0.0207 \\
								 &  & & (0.002,0.031) &  & & ($3.19\times 10^{-4}$,0.049) \\ \hline
		\end{tabular}
\end{scriptsize}\label{tab:EURO_param} 
\end{table}

Table~\ref{tab:EURO_param} and Figure~\ref{fig:EURO_post_density_box} shows that unlike the previous examples, the estimates of $\sigma,\; \phi$ and $\rho$ show different patterns. In this case, estimated $\sigma$ is approximately 10 times higher in M3.1 and M3.3 than the same under the other models. Interestingly, $\alpha$ estimates for M3.1 and M3.3 are very large in magnitude. The volatility clustering $\phi$ is estimated to be comparatively low under $h_{t+1}$ based models (0.44 for M3.1 and 0.46 for M3.3 respectively) whereas $h_t$ based models show quite high volatility clustering. One more interesting fact that may be noticed from the 95\% credible interval of $\rho$ is that it is insignificant under two out of three $h_t$-based models whereas significant under $h_{t+1}$-based models. Another important point to note is that the extremely good approximations of lead-lag correlations in M3.1 and M3.3 can perhaps be attributed to large values of $\sigma$.

\newpage{}
.

\subsection{S\&P 500 data}

{Time varying volatility of S\&P500 has been widely discussed in the literature (e.g., \citeasnoun{Eraker2003}, \citeasnoun{Bollerslev2006}, \citeasnoun{Abanto-Valle20102883}). Here we consider 1509 S\&P500 returns during January 01, 2010 and December 31, 2015, and apply the above six models on the data. Figure~\ref{fig:SNP_summary} summarizes the data.
\begin{figure}[h!]\centering
  \subfigure[Return series]{\includegraphics[width=2.5in]{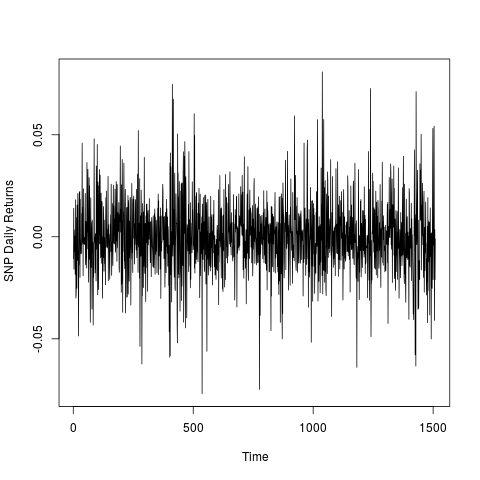}}
	\subfigure[Density plot]{\includegraphics[width=2.5in]{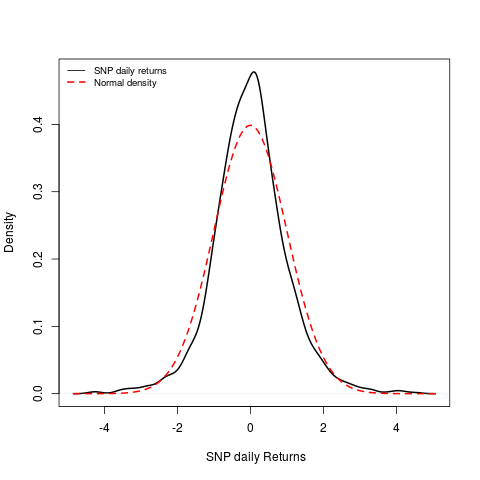}}
	\subfigure[Empirical lead-lag correlation]{\includegraphics[width=2.5in]{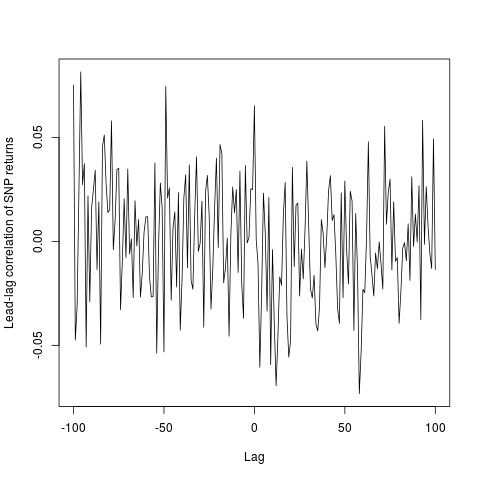}}
	\subfigure[Normal QQ-plot]{\includegraphics[width=2.5in]{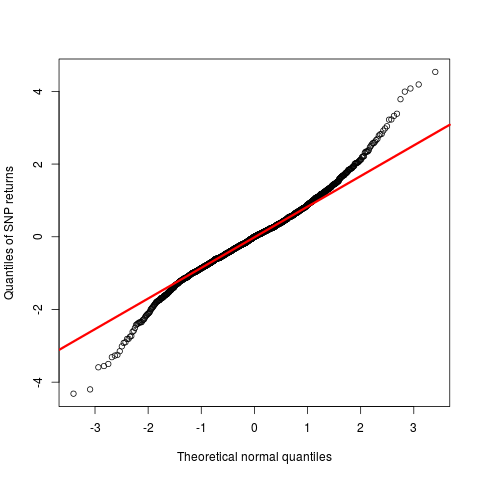}}
	\caption{Exploratory plots of daily SNP return between 1/1/2010 -- 31/12/2015.}\label{fig:SNP_summary}
\end{figure}

This is an interesting dataset with a variety of findings / results reported by experts. For instance, \citeasnoun{Schwert1987} and \citeasnoun{campbell1992} finds positive relation between market index return and its volatility, whereas \citeasnoun{GJR1993} finds the relation to be negatively correlated. The zero correlation between return and its volatility has also not been ruled out (see \citeasnoun{Wu2000}). In our implementation, we used a flat prior over $(-1,1)$ to avoid any bias.

The summary statistics of the return series (standard deviation, $0.178 \times 10^{-1}$, skewness, $Sk = -0.1095$, and kurtosis, $\kappa = 4.8811$) indicate slightly negatively skewed data with heavy tails (less heavier as compared to the previous data). The lead-lag correlations in Figure~\ref{fig:SNP_summary}(c) does not exhibit any clear trend (unlike the previous examples). The largest value (in magnitude) is at lag zero and slightly positive, $\rho_0=0.065$, unlike other examples. Figure~\ref{fig:SNP_post_leadlag} shows the model-wise comparison of the empirical and model estimated correlations.
\begin{figure}[h!]\centering
    \includegraphics[width=2.55in]{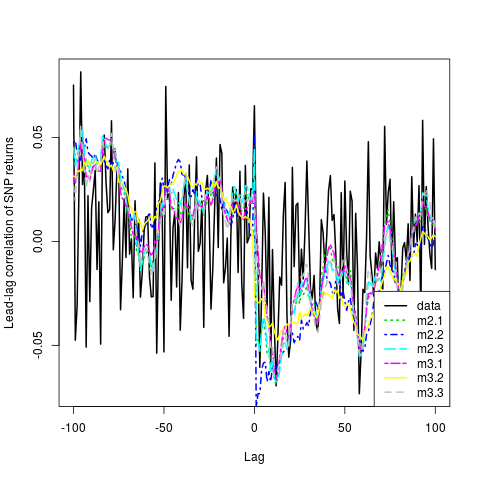}
	\includegraphics[width=2.55in]{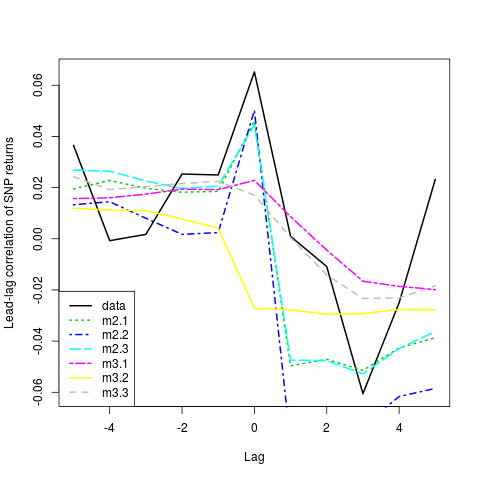}
     \caption{SNP: Comparison of model estimated lead-lag correlations, $Corr(r_t,e^{h_{t\pm k}}|\Theta, {\bf r})$ using MCMC realizations of $h_{t\pm k}$, and the empirical values, $Corr(r_t, r_t^2)$, for all six SVMs.}\label{fig:SNP_post_leadlag}
\end{figure}

Figure~\ref{fig:SNP_post_leadlag} shows that all models, except M3.2, gives positive contemporaneous correlations, and there is no clear best model. However, M2.3 and M3.1 seems to capture most of the empirical data trend.

The parameter estimates are summarized in Table~\ref{tab:SNP_param} and Figure~\ref{fig:SNP_post_density_box}.

\begin{table}[h!]\centering	
\caption{SNP: Posterior estimates of model parameters for both $h_t$-and $h_{t+1}$-based SVMs}
\begin{scriptsize}	
		\begin{tabular}{|c|c|c|c||c|c|c|}
			\hline Parameters & M2.1 & M2.2 & M2.3 & M3.1 & M3.2 & M3.3 \\ \hline
			$\alpha$   & -8.15  & -7.4083 & -8.19 & -8.23 & -7.64 & -8.1968 \\ 
								 & (-8.33,-7.96) & (-7.9995,-6.83) & (-8.37,-8.00)  & (-8.39,-8.06) & (-7.99,-7.31) & (-8.375,-8.016) \\ \hline
			$\sigma$ & 0.2 & 0.1034 & 0.1652 & 0.2236 & 0.1069 & 0.19 \\
								 & (0.12,0.28) & (0.068,0.1448) & (0.005,0.25) & (0.13,0.34) & (0.067,0.15) & (0.007,0.31) \\ \hline
			$\phi$ & 0.93 & 0.9579 & 0.92 & 0.91 & 0.9675 & 0.91 \\
								 & (0.87,0.98) & (0.9264,0.99) & (0.86,0.9738) & (0.83,0.97) & (0.94,0.99) & (0.8268,0.9749) \\ \hline
			$\rho$ & -0.17 & -0.3785 & -0.2057 & -0.014  & -0.199 & $3.5\times10^{-5}$ \\
							& (-0.38,0.035) & (-0.6357,-0.1355) & (-0.51,0.1021) & (-0.1956,0.16) & (-0.518,0.119) & (-0.002,0.021) \\ \hline
			$\nu$ & -- & 6.75 & -- & --  & 6.44 & -- \\
								 &  & (4.58,9.2) &  &   & (4.51,9.06) &  \\ \hline
			$\lambda$ & -- & -0.005 & -- & -- & -0.004 & -- \\
								 &   &  (-0.029,-0.002) &  &  &  (-0.06,0.05) &  \\ \hline
			$\pi_1$ & -- & -- & 0.0017 & --  & -- & 0.0016 \\
								 &  & & ($3.6\times 10^{-5}$,0.0039) &  & & ($3.48\times 10^{-5}$,0.0037) \\ \hline
			$\pi_2$ & -- & -- & 0.019 & --  & -- & 0.018 \\
								 &  & & ($8.5\times 10^{-4}$,0.04) &  & & ($7.6\times 10^{-4}$,0.041) \\ \hline
		\end{tabular}
\end{scriptsize}\label{tab:SNP_param} 
\end{table}

\begin{figure}[h!]\centering   
	\subfigure[$\alpha$]{\includegraphics[width=2.65in]{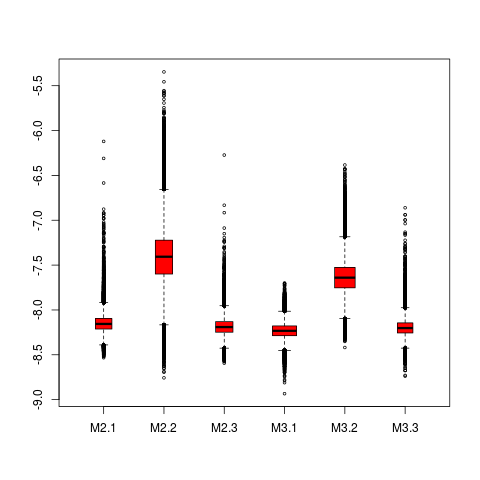}}
	\subfigure[$\sigma$]{\includegraphics[width=2.65in]{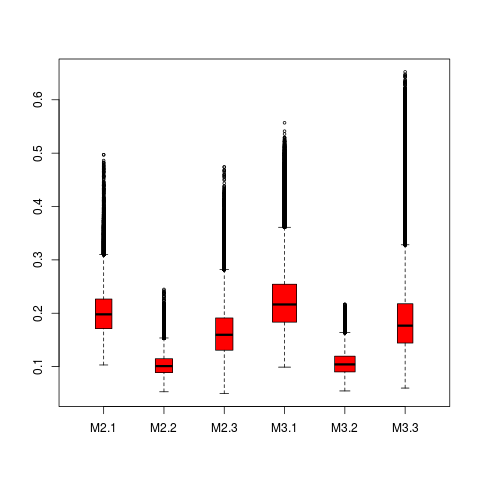}} 	
	\subfigure[$\log(\phi/(1-\phi))$]{\includegraphics[width=2.65in]{SNP_sigma_all_models.png}}
	\subfigure[$\rho$]{\includegraphics[width=2.65in]{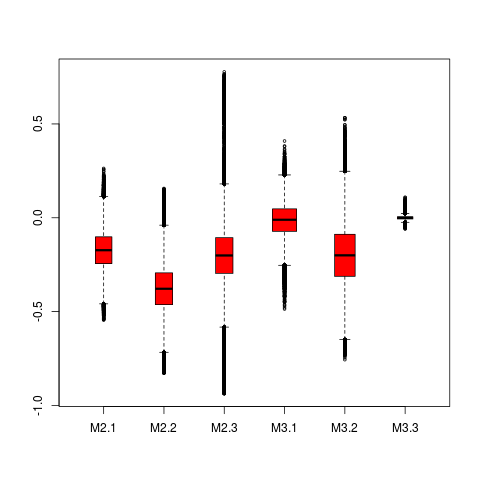}} 
\caption{SNP data: posterior distribution of $\alpha,\sigma,\phi,\rho$ based on the MCMC realizations for all models in Sections 2 and 4.} \label{fig:SNP_post_density_box}
\end{figure}

 As in earlier examples, the common parameters $\alpha, \sigma, \phi$ and $\rho$ estimates are comparable between the two classes of models (i.e., $M2.j$ vs. $M3.j$), except the values of $\rho$ in $M3.3$.
 
\subsection{Results Discussion}

We now discuss the overall findings of all four examples applied to all six SVMs. Though there are several interesting observations, a few important remarks are as follows:

(1) All tables of parameter estimates and boxplots suggest that $\rho$ is largest (in magnitude) when the return distribution is assumed to be skewed-$t$, i.e., $\rho(Mi.2)>\rho(Mi.1), \rho(Mi.3)$, for both $i=2$ and $3$. This perhaps implies that if the skewness is disentangled then the correlation between $\varepsilon_t$ and $\eta_t$, or equivalently, the leverage effect, becomes prominent.

(2) Three out of four examples (ORB price, Euro-USD rate and S\&P500 index) suggest that the estimate of $\sigma$ is smallest in skewed-$t$ models,  i.e., $\sigma(Mi.2)<\sigma(Mi.1), \sigma(Mi.3)$, for both $i=2$ and $3$.

(3) The noticeable different posterior distribution of $\rho$ for M3.3 (in all examples) can perhaps be justified as follows. Since the estimated jump probability ($\pi_1$) in the return processes of all examples are very small, let $\pi_1$ to be identically equal to zero. Further suppose a volatility jump of size $\Delta$ occurs at time $t$, i.e., $J_{2,t}=1$ and $J_{2,t-1}=0$. Then, the current return as per M3.3 at this jump time $t$ is
$$r_t = \exp\left\{\frac{\Delta}{2}\right\}\exp\left\{\frac{\alpha+\phi(h_{t-1}-\alpha)+\sigma \eta_t }{2}\right\}\varepsilon_t.$$
Thus, a part of the leverage (or the return-volatility balance) is controlled by the $\Delta$ term and the remaining part is captured by the correlation $\rho$. In other words, a smaller value of $|\rho|$ (as compared to other models) is somewhat expected. For M2.3, however, the current return will be
$$r_t = \exp\left\{\frac{\alpha+\phi(h_{t-1}-\alpha)+\sigma \eta_{t-1} }{2}\right\}\varepsilon_t,$$
where $h_t = K_{2,t-1}J_{2,t-1} + \alpha + \phi(h_{t-1}-\alpha) + \sigma\eta_{t-1}$ and $J_{2,t-1}=0$ (as per our assumption of the occurrence of a jump at time $t$). Therefore the term containing $\Delta$ does not appear in the return expression for M2.3, and there is no confounding effect.
%

\section{Conclusion} \label{sec:disc}

Modelling and analysis of the time varying volatility of returns of the risky assets has been a topic of interest for decades. Researchers have proposed a plethora of continuous-time stochastic volatility models, however, for the discrete-time setup, more innovative endeavours are still required. Among others, \citeasnoun{Ghysels1996119}, \citeasnoun{Jacquier2004185}, and \citeasnoun{Dipak2015} present a few popular discrete-time models, which we refer to as $h_{t+1}$-based SVMs (see Section~2). It turned out that \citeasnoun{Taylor1982} proposed an SVM much earlier (which we call $h_t$-based SVM in Section~3), but the naive generalization \cite{Jacquier2004185} that could accommodate more complex and realistic market phenomena violated a necessary condition called \emph{efficient market hypothesis} (EMH) which prevents arbitrage opportunities. As a result this class of $h_{t+1}$-based SVMs gained more popularity than the  $h_t$-based SVM. 

The main idea of this paper is motivated by \citeasnoun{Jacquier2004185} which attempts to generalize the model by \citeasnoun{Taylor1982}, but as \citeasnoun{Yu2005165} pointed out this model violated EMH. Recently, \citeasnoun{Mukhoti2016} revisited the results and presented a new mean-corrected model (M3.1) to make the SVM usable. In this paper, we extended this work and developed generalized $h_t$-based SVMs with correlated errors, skewed-$t$ return distribution, and jumps in the return and log-volatility processes. We also derived closed form expressions for the variance, skewness and kurtosis of the marginal return distribution, and lead-lag correlations between $r_t$ and $e^{h_{t\pm k}}$.

While comparing the two classes of SVMs, we discovered that the $h_{t+1}$-based SVMs have features that may not be desirable. For instance, with respect to the marginal return distribution, (a) the skewness measure is zero under M2.1 and M2.3, (b) important summary statistics like variance, skewness and kurtosis are free from $\rho=Corr(\varepsilon_t, \eta_t)$, and (c) contemporaneous and lagged-correlations are zero, i.e., $Corr(r_t, e^{h_{t-k}})$ are zero for $k\ge0$.

We implemented both classes of models (i.e., all six SVMs) to a variety of real-life applications (Oil price, CITI bank price, Euro-USD exchange rate, and S\&P500 index) and found some interesting features. For example, (a) the exchange rate showed slightly different estimation pattern than the others, (b) posterior distribution of the parameters are mostly similar across all other examples, except $\rho$ for M3.3, (c) overall, M3.1 ($h_t$-based base model) and M3.3 ($h_t$-based jump model) appear to be the better than other competitors in terms of estimating lead-lag correlations, (d) kurtosis estimate under $h_{t+1}$-based models are greater than the corresponding estimate under the $h_t$-based SVM.

There are several interesting unanswered questions that can be taken up as immediate future research. For instance, time-varying skewness can perhaps be modelled via $\lambda$ as a function of $t$, and instead of using $AR(1)$ as log-volatility process, one can explore more general modeling options. Of course, one can easily combine the jump models with skewed-$t$ errors, investigate correlated jumps and perhaps common jump components in such SVMs.

\bibliographystyle{ECA_jasa}
\bibliography{SVMRef}

\end{document}